\documentclass[journal]{elsarticle}
\journal{Journal of Systems and Software}
\usepackage{tabu}
\usepackage[T1]{fontenc}
\usepackage{afterpage}
\usepackage[english]{babel}
\usepackage[utf8]{inputenc}
\usepackage{float}
\usepackage{comment}
\usepackage{graphicx}
\usepackage{changepage}
\usepackage{multirow}
\usepackage{longtable}
\usepackage{url}
\usepackage{xcolor}
\usepackage{tcolorbox}
\usepackage{blindtext}
\usepackage{fontawesome}
\usepackage{amssymb}
\PassOptionsToPackage{normalem}{ulem}
\usepackage{ulem}
\usepackage{array}
\usepackage{booktabs}
\usepackage{amsmath}
\usepackage{comment}
\usepackage{pgfplots}
\usepackage{pgf-pie}
\usepackage{tabulary}
\usepackage{etoolbox}
\usepackage{listings}
\usepackage{xcolor}
\usepackage{pdflscape}
\usepackage{enumitem}
\usepackage{enumitem}
\usepackage{adjustbox}
\usepackage{rotating}
\usepackage{tablefootnote}
\usepackage{subcaption}
\usepackage{pgfplotstable}
\usepackage{tikz}
\usepackage{mdframed}
\usepackage[nottoc]{tocbibind}
\usepackage{vcell}
\usepackage{colortbl}
\usepackage{graphics}
\usepackage{enumitem}
\usepackage{tabularx}
\usepackage{pdflscape,array,booktabs, multirow}
\usepackage{lipsum}
\usepackage{blindtext}
\usepackage{makecell}
\usepackage{ragged2e}
\usepackage{ltablex}
\usepackage{indentfirst}
\newcolumntype{C}{>{\Centering}X}
\newcolumntype{L}{>{\RaggedRight}X}
\newcolumntype{R}{>{\RaggedLeft}X}
\newcolumntype{P}[1]{>{\centering\arraybackslash}p{#1}}
\usepackage{hyperref}
\hypersetup
{
  colorlinks   = true, 
  urlcolor     = red, 
  linkcolor    = blue, 
  citecolor    = blue 
}
\usepackage{geometry}
\pgfplotsset{compat=1.9}

\usepackage{lineno}
\usepackage{etoolbox}

\begin{document}
\raggedbottom
\setlength{\textfloatsep}{10pt plus 1.0pt minus 2.0pt}

\title{Software Architecture for Quantum Computing Systems - A Systematic Review}
\author[1]{Arif Ali Khan\corref{mycorrespondingauthor}}
\ead{arif.khan@oulu.fi}
\address[1]{M3S Empirical Software Engineering Research
Unit, University of Oulu, 90014 Oulu, Finland}
\author[2]{Aakash Ahmad}
\ead{a.ahmad13@lancaster.ac.uk}
\address[2]{School of Computing and Communications, Lancaster University Leipzig, Germany}

\author[3]{Muhammad Waseem}
\address[3]{School of Computer Science, Wuhan University, Wuhan, China}
\cortext[mycorrespondingauthor]{Corresponding authors}
\ead{m.waseem@whu.edu.cn}

\author[3]{Peng Liang}
\ead{liangp@whu.edu.cn}
\author[4]{Mahdi Fahmideh}
\ead{mahdi.fahmideh@usq.edu.au}
\address[4]{School of Business at University of Southern Queensland, Queensland, Australia}
\author[5]{Tommi Mikkonen}
\ead{tommi.j.mikkonen@jyu.fi}
\address[5]{Faculty of Information Technology, University of Jyväskylä, FI-40014 Jyvaskyla, Finland}
\author[5]{Pekka Abrahamsson}
\ead{pekka.abrahamsson@jyu.fi}

\begin{abstract}
Quantum computing systems rely on the principles of quantum mechanics to perform a multitude of computationally challenging tasks more efficiently than their classical counterparts. The architecture of software-intensive systems can empower architects who can leverage architecture-centric processes, practices, description languages to model, develop, and evolve quantum computing software (quantum software for short) at higher abstraction levels. We conducted a Systematic Literature Review (SLR) to investigate (i) architectural process, (ii) modeling notations, (iii) architecture design patterns, (iv) tool support, and (iv) challenging factors for quantum software architecture. Results of the SLR indicate that quantum software represents a new genre of software-intensive systems; however, existing processes and notations can be tailored to derive the architecting activities and develop modeling languages for quantum software. Quantum bits (Qubits) mapped to Quantum gates (Qugates) can be represented as architectural components and connectors that implement quantum software. Tool-chains can incorporate reusable knowledge and human roles (e.g., quantum domain engineers, quantum code developers) to automate and customize the architectural process. Results of this SLR can facilitate researchers and practitioners to develop new hypotheses to be tested, derive reference architectures, and leverage architecture-centric principles and practices to engineer emerging and next generations of quantum software.
\end{abstract}
\maketitle

\section{Introduction}

Quantum computing relies on quantum mechanics, a discipline more familiar and center of attention to physicists rather than computer scientists or software engineers \cite{zhao2020quantum}\cite{deutsch1985quantum}\cite{dirac1981principles}. However, in recent years, with an emergence of quantum algorithms and Quantum Programming Languages (QPL), software programmers have been able to exploit the theory and principle of quantum mechanics to process information and perform specific computation tasks faster than classical computing systems \cite{chong2017programming}\cite{ying2016foundations}. Compared to classical algorithms for computation, quantum algorithms have the potential to solve a set of computationally challenging problems such as nature-inspired computing, financial modeling, and advanced encryption with increased efficiency  \cite{montanaro2016quantum}\cite{grimsley2019adaptive}\cite{kruger2020quantum}. Quantum computing attributes (e.g., Qubits, superposition, entanglement, and interference) lie at the heart of quantum information processing \cite{zeilinger1999experiment}\cite{gay2006quantum}. Quantum programming languages that implement quantum algorithms enable quantum supremacy in computing that is lacking in traditional computing systems \cite{montanaro2016quantum}\cite{gay2006quantum}\cite{sofge2008survey}. One class of such problems relate to information and computation science that requires large amounts of parallel processing \cite{nguyen2022software} for tackling challenges, such as optimization, encryption, big data analytics, and machine learning \cite{biamonte2017quantum}\cite{rebentrost2014quantum}. Other set of problems relate to efficient and accurate simulation of quantum systems in natural sciences, such as physics \cite{grimsley2019adaptive}, chemistry \cite{mcardle2020quantum}, mathematics \cite{kruger2020quantum}, and challenges relating to their applications \cite{stepney2005journeys}\cite{childs2018toward}\cite{mosca2018cybersecurity}. However, QPL and their underlying algorithms focus on computation and implementation details to produce executable specifications, but lack an overall global view of the software systems under design. Source code based implementation details undermine architectural view(s) as system blueprint, that can compromise the quality and functionality of end product, i.e., quantum software \cite{zhao2020quantum}\cite{moguel2020roadmap}\cite{piattini2021toward}. Technology giants are scaling up their financial and strategic investments in quantum computing platforms, more specifically quantum programming languages such as Q\# from Microsoft, Qiskit from IBM, and Cirq from Google, however; quantum software engineering and development is still in its infancy \cite{WinNTMS}\cite{behera2019designing}\cite{courtland2017google}. Some recent research studies also indicate that quantum software projects that overlook design principles to primarily focus on quantum source code implementations, often lead to faulty implementations and bugs in quantum software \cite{zhao2021bugs4q}\cite{Qubug2}.

Software architecture as described in the ISO/IEC 42010 standard provides a global view of software-intensive systems, representing their blue-print, by abstracting complex implementation details with architectural components and connectors \cite{WinNTISO}\cite{RQ1-1}\cite{RQ1-2}. Software developers and architects have successfully used architectural descriptions and specifications to design, develop, validate, and evolve software-intensive system at higher-level of abstractions while maintaining system functionality and quality \cite{RQ1-5}\cite{RQ1-6}. Architectural models have been exploited to design, develop, and validate emerging generations of software-intensive systems including but not limited to the internet of things, blockchain applications, and artificially intelligent systems \cite{alreshidi2019architecting}\cite{IoTTSE}\cite{xu2017taxonomy}\cite{CSURBC}\cite{graef2021software}. Quantum Software Architecture (QSA), as a new genre of Software Architectures (SA), can provide architectural descriptions (i.e., components, connectors, and configurations) to design and develop quantum software, while abstracting complex and implementation specific tasks \cite{RQ1-1}\cite{RQ1-10}. Specifically, architectural components can represent modules of source code while architectural connectors specify interactions between modules to represent the structure and behavior of a system \cite{RQ1-10}. Transformation from abstract high-level models (i.e., design artifacts) to low-level executable specifications (i.e., source code artifacts) can be enabled via model-driven architecting of quantum software \cite{RQ1-11}\cite{perez2021modelling}. However, QSA as an emerging discipline remains an under-explored area by the current generation of designers and architects who find themselves less prepared to tackle the challenges related to QSA in the development life-cycle of quantum software \cite{WinNTRoberto}\cite{jeff2021}\cite{WinNTYehuda}. Despite a plethora of published research in recent years that focuses on engineering and architecting quantum software, there do not exist any evidence, i.e., empirical study or data-driven analysis to consolidate a collective impact of existing research on architecting quantum software \cite{RQ1-10}\cite{RQ1-11}\cite{perez2021modelling}\cite{WinNTRoberto}.

Systematic Literature Reviews (SLRs) rely on Evidence-based Software Engineering (EBSE) approach to identify, classify, compare, and synthesise published research as an evidence to empirically investigate the topic under investigation \cite{RQ1-6}\cite{keele2007guidelines}. Recently, a number of SLRs and review based studies have been conducted to investigate the application of Software Engineering (SE) to quantum computing systems, however; there is no effort to review the state-of-the-art on architecting quantum software \cite{zhao2020quantum}\cite{moguel2020roadmap}\cite{piattini2021toward}\cite{gill2020quantum}. Therefore, the objective of this review is to complement SE based studies and specifically focus on \textit{identification, classification, and synthesis of the published research on the role that software architecture plays in developing quantum computing systems}. We aim to investigate the core concepts, underpinning fundamentals of software architectural aspects, often overlooked in SE focused studies, by outlining a number of Research Questions (RQs). These RQs focus on (i) architectural process (unifying architecting activities), (ii) modeling notations (architectural representation), (iii) patterns and design decisions (reusable knowledge and best practices), (iv) tool support (enabling automation and customisation) and (v) emerging challenges for quantum software architectures. These RQs are motivated by academic research and industrial studies on software architecture that highlight the needs for process-centric architecting, where a process acts as an umbrella to support various architectural aspects  \cite{RQ1-1}\cite{RQ1-5}. Moreover, in quantum software engineering lifecycle \cite{dey2020qdlc}, during system design, architectural aspects such as software modeling, patterns, tools, and human roles are as fundamental for architecture-centric engineering of quantum software \cite{piattini2021toward}. Results and findings of this SLR complement existing surveys on Quantum Software Engineering (QSE) and can provide foundations for further secondary studies that can explore architectural principles and practices to design and develop quantum software.

The results of this SLR indicate that although quantum software represents a new generation of software applications, foundations for quantum software architectures are grounded in architectural processes and architecting activities of classical systems (e.g., object, service, or component-based) \cite{kruger2020quantum}\cite{RQ1-5}\cite{RQ1-6}\cite{diadamo2021qunetsim}. Quantum-specific features involving Qubits (e.g., quantum entanglement and quantum superposition) elaborated later, do require tailored architectural processes and modeling notations, such as exploiting the Unified Modeling Language (UML) to effectively address the challenges of the quantum age architectures \cite{perez2021modelling}. Specifically, existing processes and notations need customisation to enable co-design of quantum systems that can enable the mapping between Qugates and Qubits to software architectural components and connectors. Tool-chain to support quantum architecting process can facilitate system and software architects to achieve automation and incorporate human decision support while designing and implementing quantum software. The results of the SLR can be beneficial for:
\begin{enumerate}[label=(\roman*)]
 \item Researchers who are interested in understanding theory and principles of architecture-intensive development, establishing new hypotheses to be tested, and developing reference architectures and solutions for quantum software. 
 \item Practitioners who would like to understand the architecting activities, patterns as reusable knowledge, existing and required tool chain, and the extent to which the academic research can be leveraged to develop industry scale solutions for quantum software.
\end{enumerate}

The rest of the paper is organized as follows: Section \ref{Context} presents the context and background of this research study. Section \ref{research methodology} details the research methodology to conduct the study. Section \ref{demographyresults} - Section \ref{RQ2} present the result of the study. Section \ref{keyfindings} discusses the core finding and implications of the study results. Section \ref{threats to validity} elaborated on threats to the validity of the research. Section \ref{related work} reviews and provides comparative analysis of the most relevant research studies. Section \ref{conclusions} concludes the study with a discussion of potential future research.






\section{Context: Architecting Software for Quantum Computing} \label{Context}
This section contextualises quantum computing systems in terms of their building blocks, i.e., (a) quantum hardware, (b) quantum software, and (c) quantum software architecture as shown in Figure \ref{QC}. More specifically, Figure \ref{QC} provides a visual reference that correlates the Qubits and Qugates to quantum source code, representing design and implementation phase of QSE life-cycle. Software architectural components and connectors provide a blue-print to implement the quantum source code. We use the illustrations in Figure \ref{QC}, elaborated below, to introduce fundamental concepts and terminologies that will be used throughout the paper.

\subsection{Quantum Computing Systems}

\begin{figure}[h!]
 \centering
 \includegraphics[scale=0.7]{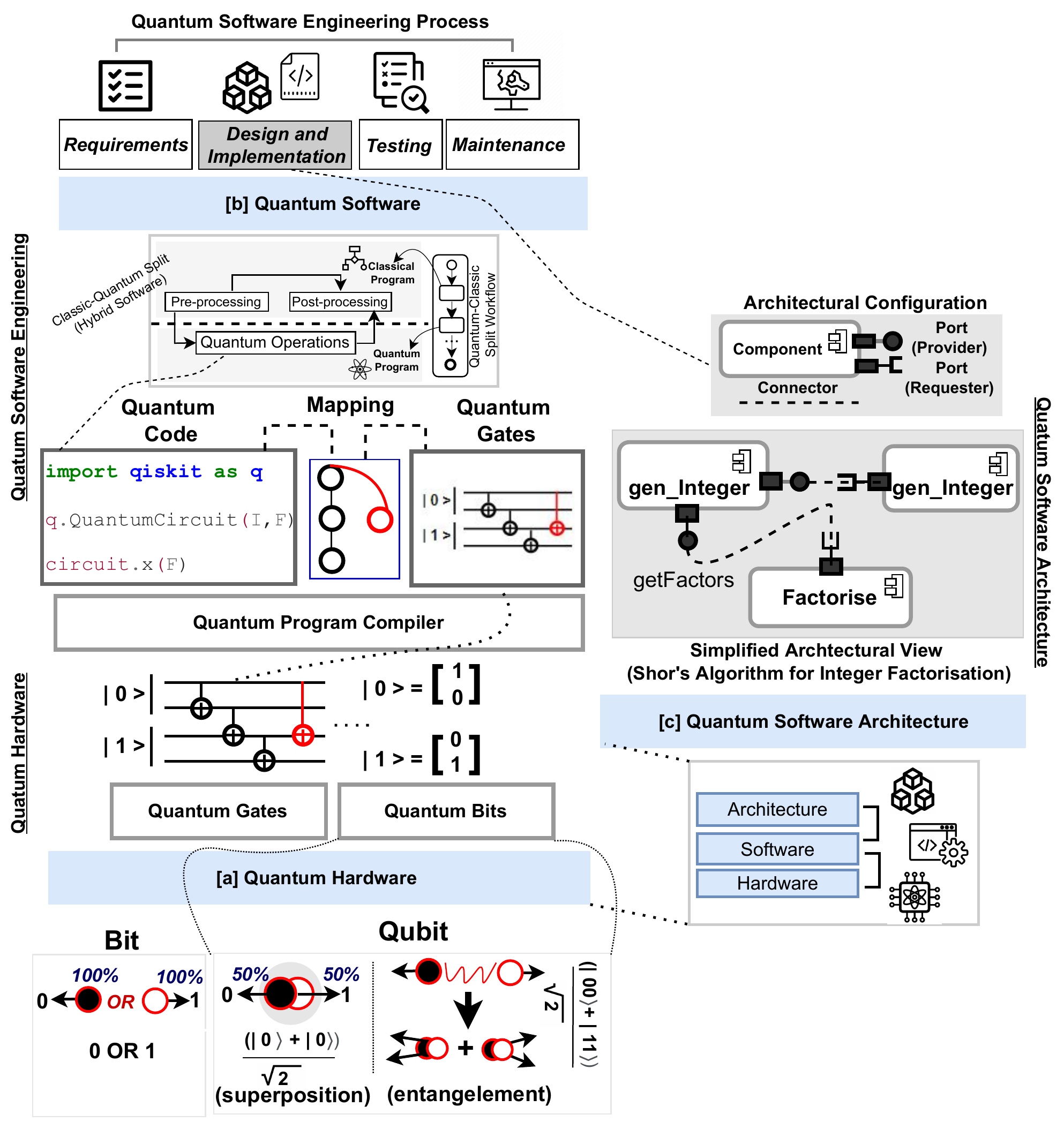} 
 	\caption{A Simplified View of Quantum Computing Systems ([a] Quantum Hardware, [b] Quantum Software, [c] Quantum Software Architecture)}
	\label{QC}
\end{figure}

To gain strategic advantages of quantum information processing, technology giants, such as IBM, Google, Microsoft and governmental organizations are heavily investing in the research and development of quantum  systems \cite{WinNTMS}\cite{behera2019designing}\cite{courtland2017google}\cite{TopCountries}.
From the system’s engineering perspective, as shown in Figure \ref{QC}, fundamental to quantum computing hardware is the concept of Qubit (quantum bit) that represents the most fundamental unit of quantum information processing \cite{zeilinger1999experiment}\cite{gay2006quantum}. Contrary to the classical bit (binary digit) that is expressed as [1, 0] in digital computing systems, a Qubit represents a two-state quantum computer and these two states are specified as $|0\rangle$ and $|1\rangle$. The combinations of bits represent flow of digital information that alters the state of binary logic gates (on: 0 off: 1) to make digital systems work. Analogous to the binary gates, quantum gate (a.k.a. the quantum logic) represents the building blocks of a quantum circuit and transits its state via Qubit \cite{gay2006quantum} as in Equation (\ref{EQ-1}). A Qubit can be in a state  $|0\rangle = \begin{bmatrix} 1 \\ 0 \end{bmatrix}$ and $|1\rangle = \begin{bmatrix} 0 \\ 1 \end{bmatrix}$ or (unlike a classical bit) in a linear combination of both states. 

\begin{equation}\label{EQ-1}
 |0\rangle  =  \left[ \begin{array}{c} 1 \\ 0 \end{array} \right] ~~~~~ + ~~~~~  |1\rangle  =  \left[ \begin{array}{c} 0 \\ 1 \end{array} \right]   
\end{equation}

In Figure \ref{QC} [a], we illustrate and elaborate on the distinction between a Bit and Qubit. A Bit is like a gate in an electronic circuit that can be either on or off, whereas a Qubit uses the unique properties of quantum mechanics to provide a unit that can be one or zero - or anything in between. The bit can take a value of `0' or `1' as either `Off' or `On' with 100\% probability (left). A qubit can be in a state of |0⟩ or |1⟩ or in a superposition state with 50\% |0$\rangle$ and 50\% |1$\rangle$, superposition state (left).  Two Qubits are in an entangled state (right) - entangled qubits are linked such that by looking (i.e., measuring) one of these two, will reveal the state of other Qubit. Further details about Qubit and Qugate in the context of operationalising the QC systems can be found in \cite{zhao2020quantum} \cite{zeilinger1999experiment}. Like the classical computing systems, controlling the Qubits that manipulate Qugates, there is a need for quantum software systems and applications that can exploit benefits of quantum information processing by operationalising quantum computers. For example, QuNetSim \cite{diadamo2021qunetsim} is a Python software framework that is capable of managing quantum circuits to simulate processing and transmission of quantum information via quantum networks. Figure \ref{QC} shows that in order to enable quantum software applications to utilise quantum hardware, there is a need for quantum code compilers that can translate high-level computational instructions into machine translated code to control quantum hardware \cite{chong2017programming}\cite{garhwal2021quantum}. As a typical example of such compilation are the solutions by proposed by Ying \cite{ying2016foundations} and, Krüger and Mauerer \cite{kruger2020quantum}, which receive the compiled code that can be executed or simulated on quantum platforms to enable quantum processing for optimising solutions regarding unstructured data searching, parallel processing, and nature inspired computing. In recent years, a plethora of research and development has emerged that focused on quantum algorithms and programming languages to address the above-mentioned computational challenges effectively and efficiently \cite{garhwal2021quantum}. Quantum algorithms have the potential to provide computation efficiency to software engineering problems in areas including but not limited to data mining, machine learning, and cryptography that do not scale optimally on non-quantum computing platforms \cite{QuantumProspects}. Despite the significance of quantum programming languages to produce executable specifications for quantum hardware; there is a need for overall engineering lifecycle(s) that goes beyond level of source code to specify, execute, validate, and evolve software-intensive system based on required functionality and desired quality \cite{moguel2020roadmap}\cite{piattini2021toward}.

\subsection{Software Engineering (SE) for Quantum Computing}
Software engineering, as defined in the ISO/IEC/IEEE 90003:2018 standard aims to apply engineering principles and practices to design, develop, validate, deploy, and evolve software-intensive systems effectively \cite{WinNTISO9001}. In recent years, SE focused research and development started to tackle, such as quantum software models, their algorithmic specifications, and simulated evaluations to leverage benefits of quantum hardware for quantum information processing \cite{montanaro2016quantum}\cite{grimsley2019adaptive}\cite{rebentrost2014quantum}\cite{childs2018toward}\cite{svore2006layered}. More specifically, software engineers can leverage SE practices and patterns by following software process(es) that comprises of a multitude of engineering activities including but not limited to requirements engineering, design, implementation, evaluation, and deployment, as shown in Figure \ref{QC}. SE activities adopted from quantum and classical software engineering concepts are used to represent a simplified view of quantum SE process (see Figure \ref{QC} (b)) \cite{zhao2020quantum}\cite{merlin1997SE}. Such generalised process can be tailored (adding, removing, and/or customising any activities) as per the context of system development. 

Quantum computing systems are in a phase of continuous evolution and consequently quantum SE represents a new generation of software-intensive engineering activities to develop applications that can control the underlying hardware \cite{piattini2021toward}. In recent years, research communities on software engineering and software architecture have focused on establishing dedicated forums, i.e., conferences, workshops and alike forums in an attempt to set the agenda(s), streamline emergent challenges, and propose community wide initiatives to engineer and architect quantum software \cite{QSEWorkshop}\cite{QSAWorkshop}. These QSE focused research communities intend to gather researchers and practitioners and provide a forum to collaborate and explore the possibilities to exploit existing software engineering methods that can be applied to quantum era computing and software systems \cite{QSEReport}\cite{SE-QC}. The Quantum Flagship represents a prime example to support sustainable research and development for consolidating and expanding scientific leadership, achieving excellence and innovation in quantum computing technologies \cite{QuantumFlagship}. QSE process may involve an additional challenge of managing hybrid applications and algorithms. A hybrid application and its underlying implementation involve splitting the overall application into classical modules (pre/post-processing) and quantum modules (quantum computation) referred to as the quantum-classic split  \cite{Classic-Quantum}, as shown in Figure \ref{QC} (b). Research on the quantum-classic split is gaining attention with an aim to develop QSE process(es) that enable quantum software designers and developers to engineer hybrid applications by applying the quantum-classic split pattern \cite{QMLHybrid}.

In addition to the needs for innovative technologies and processes, principle, and practices that specifically tackle challenges for quantum software modeling and architecting, coding, and simulation, existing classical SE processes can be customized to engineer and develop quantum software \cite{svore2006layered}\cite{baczewski2017co}. For example, the concept of architectural modeling as a generic architecting activity, can be customised with initiatives like quantum UML profile, exploiting the UML activity diagrams that could help model parallel computing for quantum search algorithms \cite{perez2021modelling}. UML profiles for quantum systems enable software designers to create multiple views as different perspectives of system under design. For example, the designer can utilise the activity diagram to design quantum circuits \cite{perez2021modelling} or utilise use case, sequence, or deployment diagrams to design the interaction, control flow, and configuration views of classical-quantum software \cite{QMLHybrid}. Similarly, existing requirements engineering process can be tailored to support requirements for quantum (i.e., quantum entanglement) that is missing in the existing models. In SE process(es), architecting represents a pivotal activity that accumulates system requirements as a model thus leading to software implementation, validation, and evolution while maintaining a global view of the system and managing architectural trade-offs \cite{RQ1-5}\cite{RQ1-6}.

\subsection{Architecture for Quantum Software}
Architecture of software intensive systems, as described in the ISO/IEC/IEEE 42010:2022 standard, aims to abstract complex and implementation specific details to represent system blueprint in an implementation and technology neutral way \cite{WinNTISO}. Empirically-grounded academic research and industrial studies on architecting software-intensive systems have highlighted that there is no unified view to represent software architectures \cite{RQ1-1}\cite {RQ1-5}\cite {TSEADL}. Different architectural views (also referred to as architectural models or representations) can also be attributed to a multitude of modeling approaches supported via UML, ADL, and graph models that allow software practitioners to create customised architectural view(s) that fits their context in a specific architecting activity \cite{QMLHybrid}\cite{UML}. For example, considering the 4 + 1 architectural view \cite{RQ1-1}, requirements engineers may be more interested in the interaction model(s) expressed as graphs or UML use case diagrams that capture architecturally significant requirements (functionality and quality of system). In comparison, software developers and quality engineers/testers are more likely to utilise the component and connector models that represent modules of source code and their interactions, and runtime view that models system execution as UML sequence diagrams. As per the 4 + 1 architectural view, in this study, we have mainly relied on the component and connector architecture model (Figure 1) that represents software in terms of computations and data stores. However, during architectural review and syntheis, the component and connector architectural models alone are not sufficient and the effort to consolidate a singular or unified view that supports various architectural activities may be impractical. Once expressed, some architectural models, i.e., model-driven architecture can help to generate the necessary skeleton or libraries of source code in a (semi-) automated way using model-driven engineering \cite{RQ1-11}. In recent years, architectural models and notations have proven to be successful to design and develop software intensive systems by enabling reusability (patterns and styles), evolvability (architectural reconfigurations), and elasticity (auto-scaling) \cite{RQ1-1}\cite{RQ1-2}. Figure \ref{QC} (c) illustrates a partial architectural view of a quantum algorithm to factorise integers that is modeled as UML component diagram \cite{perez2021modelling}. The architectural view abstracts the source code level details to present design decisions in terms of components (Shor\_Factor, Shor\_Order) that coordinate via a connector (getFactors) for integer factorisation. Architecture in itself represents non-executable specifications of the quantum search system, however; the application of model-driven engineering can help architects and designers to derive source code directly from architecture models.

In the overall view of Figure \ref{QC}, we can conclude that in quantum computing systems, software architecture represents a blue-print to develop software systems and applications that manipulate quantum hardware. Quantum software projects primarily focused on producing quantum source code while overlooking quantum software design are often prone to bugs and unfulfilled requirements \cite{Qubug2}. The role of software architecture in quantum SE is pivotal to develop the requirements, which lead to software designing, coding, validation, and deployment, all facilitated using architectural notations. Software architecture for quantum computing systems (quantum software architecture) can empower the role of software engineers and developers to create models that act as basis for system implementation. Based on the architectural models, model driven engineering and development can be exploited for the automated generation of quantum source code (code modules and their interactions) from the corresponding quantum software architecture (based on architectural components and their connectors) \cite{ying2016foundations}\cite{RQ1-11}\cite{jeff2021}.

\section{Research Methodology} 
\label{research methodology}
\begin{figure}[h!]
 \centering
  \includegraphics[width=\linewidth]{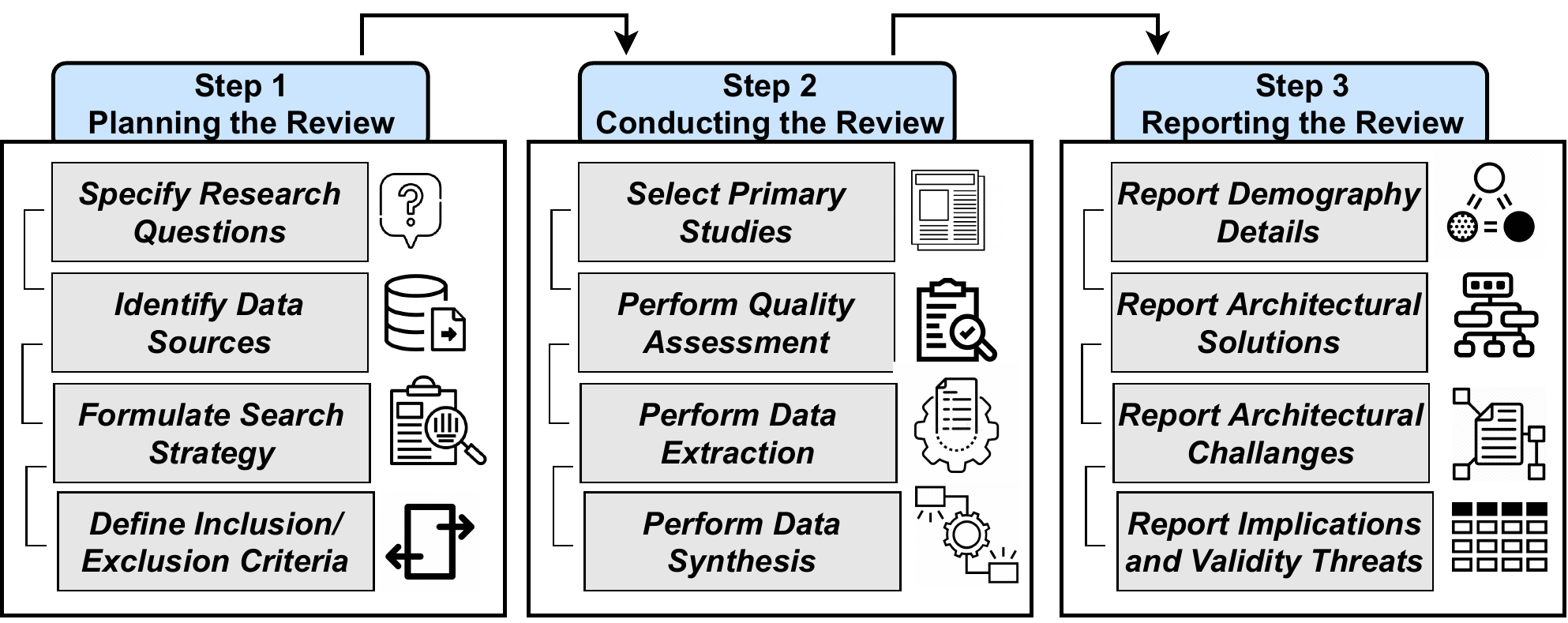}
 	\caption{An overview of the research methodology for SLR}
	\label{Fig:researchmethodologyoverview}
\end{figure}

We followed EBSE approach to conduct this research \cite{EBSE}. As part of our research methodology, we adopted the Systematic Literature Review (SLR) approach to identify, analyse, and investigate the available literature based on the outlined research questions. Specifically, SLR follows the principle of evidence-based software engineering approach to adopt a rigorous process for conducting the review based on well-defined protocol to extract, analyse, and report the results \cite{kitchenham2004evidence}. SLR provides  \textit{“a means of evaluating and interpreting all available research relevant to a particular research question, topic area, or phenomenon of interest”}  \cite{keele2007guidelines}. We followed the guidelines provided by Kitchenham and Charters to conduct this SLR \cite{keele2007guidelines}, which consists of three core steps, i.e., \textit{planning, conducting,} and \textit{reporting} the review as illustrated in Figure \ref{Fig:researchmethodologyoverview}.

Each step of SLR, as illustrated in Figure \ref{Fig:researchmethodologyoverview}, is elaborated below. While performing the literary reviews and secondary studies in the context of software engineering research, there is an ongoing debate about conducting the Multivocal Literature Reviews (MLRs) - including grey literature - instead of SLRs for fast evolving areas like quantum computing and quantum software engineering \cite{MLR}. We preferred the SLR, based on the guidelines in \cite{keele2007guidelines}, to only review peer-reviewed published research as secondary studies on quantum software architecting. Non-peer reviewed studies and grey literature are also discussed to discuss the results of primary studies, however; such studies and literature are complementary and are not included in the list of primary studies for SLR.      

\subsection{Planning the review}
As the initial step, the planning phase starts with developing the research questions that encapsulate the key research objectives of the SLR.

\subsubsection{Step 1: Specify research questions} \label{RQs}
We outline the Research Questions (RQs) to investigate multi-faceted information including \textit{demography, architectural activities, architectural modeling notations, architectural design patterns, tools and frameworks, and challenges}. The RQs to investigate the mentioned multi-faceted information are outlined and the details along rationale of each RQ is provided in Table \ref{Tab:RQsTable}. Answer to the reported RQs helps us document the SLR results described in subsequent sections of this paper.

\afterpage{%
{\renewcommand{\arraystretch}{1}\footnotesize
\begin{longtabu} to \textwidth {|c|p{0.3\textwidth}|p{0.6\textwidth}|}
\caption{Research Questions of this SLR}
\label{Tab:RQsTable}
\\ \hline
\hline
\endfirsthead
\multicolumn{3}{c}%
{\tablename\ \thetable\ -- \textit{Continued from previous page}} \\
\hline
\textbf{\#} & \textbf{Research Question} & \textbf{Rationale}  \\
\hline
\endhead
\hline \multicolumn{3}{r}{\textit{Continued on next page}} \\
\endfoot
\hline
\endlastfoot
\multicolumn{3}{|c|}{\textbf{A: Demographic details of published research}} \\ \hline
\textbf{\#}&\textbf{Research Question} & \textbf{Rationale}\\ \hline
\textbf{RQ1.1} & What are the types and the frequency of publications on quantum software architecture? &	This RQ aims to pinpoint the types of publications (e.g., journal articles, conference proceedings) and highlight the frequency of publications (number of publications per year). The RQ provides an understanding of the research progress (i.e., type and frequency published research over the years) with respect to the topic under investigation.\\\hline
\textbf{RQ1.2} & What are the research types and reported contributions in published studies on quantum software architecture?	& Types of research (i.e., solution type, evaluation type) and research contributions help us to understand the diversity of published research, solutions to address the problems, empirical foundations, and theoretical principles as the available evidence in the SLR.\\\hline
\textbf{RQ1.3}	& What are the application domains to which the proposed architectural solutions can be applied? & Application domain refers to the areas (e.g., network security, system engineering) to which architectural solutions can be applied to address specific challenges. A classification of application domains help us understand the extent to which architectural solutions address software design challenges pertaining to different areas . \\\hline
\multicolumn{3}{|c|}{\textbf{Architectural solutions for quantum software and emerging challenges}} \\ \hline
\textbf{\#}&\textbf{Research Question} & \textbf{Rationale}\\ \hline
\textbf{RQ2.1}	& Are there any architectural processes for quantum software?	& Architectural process include a number of architecting activities to provide a step-wise and incremental approach to develop architectural solutions. By investigating the architectural process and its underlying architecting activities, we can understand  architectural analysis, synthesis, and evaluation of of proposed solutions. \\\hline
\textbf{RQ2.2} & What modeling notations have been used to represent quantum software architectural solutions? &	Modeling notations visually depict the detail sequence of architecting activities and show the relations between the numerous units of the software system. Answer to this RQ will give an understanding of existing graphical notation used to specify quantum software architecture.\\\hline
\textbf{RQ2.3}& What patterns exist for quantum software architectures?& Patterns represent reusable knowledge and best practices to design and implement software solutions. The answer to this RQ will help to investigate the patterns which reveal reusable (architectural) knowledge and best practices to architect quantum software systems.\\\hline
\textbf{RQ2.4} & Are there any tools and/or frameworks to support automation and customization of architectural solutions for quantum software? &	To study the available tools and framework support that can enable automation and customization (i.e., user decision support) of the architectural process and its activities. We aim to further analyze tools that complement the architectural solutions with their automation and customisation.\\\hline

\textbf{RQ2.5} &	What challenges have been reported for quantum software architecture? &	Various challenges could impact the process of developing quantum software architecture. Analysing the challenges will pinpoint the issues and factors that impact architectural solutions for quantum software.\\\hline

\end{longtabu}
}}

\subsubsection{Step 2: Identify data sources}
In systematic reviews and mapping studies, Electronic Data Sources (EDS) allow an automated search, based on predefined and often customised search string(s), to identify the relevant literature on a topic under investigation \cite{chen2010towards}. A number of empirical studies have investigated methods for conducting systematic searches along with putting forward a list of EDS that can help select literature efficiently while minimising the potential bias and risk of missing relevant data \cite{ HybridSearch}. Based on the recommendations for adopting a systematic search process and selecting the most relevant data source, we selected five EDS for an automated search \cite{SLRSearch}. These EDS include ACM Digital Library, IEEE Xplore, Science Direct, SpringerLink, and Wiley Online Library that represent prominent sources to search literature on computing in general and software engineering and software architecture research in particular. The list of EDS that we selected is not an exhaustive, nor does it guarantee to cover all possible existing literature, however, prior empirically-based studies on SLRs have highlighted these five electronic sources as necessarily sufficient and appropriate to identify the relevant literature \cite{chen2010towards}\cite {SLRSearch}.

\subsubsection{Step 3: Formulate search strategy} \label{Search strategy}
The first three authors analyzed the RQs to identify the key terms or keywords. Moreover, all the authors were invited to participate in the group meeting to finalize the key terms. The aim of reporting the key research terms is to develop the search string and explore the selected digital libraries using that string. Finally, the authors agreed to consider the following search string for the data search:\\
\textit{(Software) AND (Architecture OR Design OR Framework OR Pattern) AND (Quantum)}\\
The key terms are concatenated using the \textit{“OR”} and \textit{“AND”} boolean operators to develop the above-given search string. The decision to finalise the search string was based on a pilot search of relevant literature on IEEE eXplore and Google Scholar. In the pilot search, we aimed at identifying the titles of existing studies and various synonyms used to refer to software architecture in the quantum computing context. For example, we observed that use of key term \textit{'model'} as a synonym for architecture yielded a significantly large but irrelevant number of studies that discuss software process models (focused on QSE rather than QSA). Based on the consensus of the researchers, we omitted the key term \textit{'model'} to avoid an exhaustive search space. Moreover, based on the pilot searching phase, we included key term \textit{'framework'} that did identify some relevant studies. The main goal of the final search string was to identify the most relevant literature as much as possible while avoiding potentially irrelevant studies that can exhaust manual scanning of titles, keywords, and abstract for study selection. The replication package based on the given search string is provided in \cite{ReplPackage}.

\subsubsection{Step 4: Define inclusion and exclusion criteria} \label{inclusion & exclusions}
Based on the guidelines by Kitchenham et al. \cite{keele2007guidelines} for including or excluding the identified studies, we outlined the inclusion and exclusion criteria in Table \ref{Tab:inclusionexclusion}. By following the criteria any irrelevant, redundant, or non-English studies were excluded. Study inclusion and exclusion was followed by a quality assessment step to assess the quality of each included study and eliminate any study that did not satisfy the qualitative assessment criteria (see Section \ref{quality assessment}). The inclusion and exclusion criteria filters the search findings returned by the search string. The key points of the criteria were developed by the first three authors based on \cite{keele2007guidelines}. Table \ref{Tab:inclusionexclusion} provides the criteria for the inclusion and exclusion of the literature for review along with the codes (Incl 1-4: as the inclusion criteria and Excl 1-4: as the exclusion criteria). We discuss the details in Table \ref{Tab:inclusionexclusion} later to elaborate the selection of primary studies to be included in the SLR.   

\subsection{Conducting the review}
The second phase of the SLR process is conducting the review, which is based on the protocol defined in the first phase, i.e., \textit{planning the review} (See Figure \ref{Fig:researchmethodologyoverview}). Following are the key steps involved in this phase:

\begin{figure}[!h]
 \centering
 \includegraphics[width=\linewidth ]{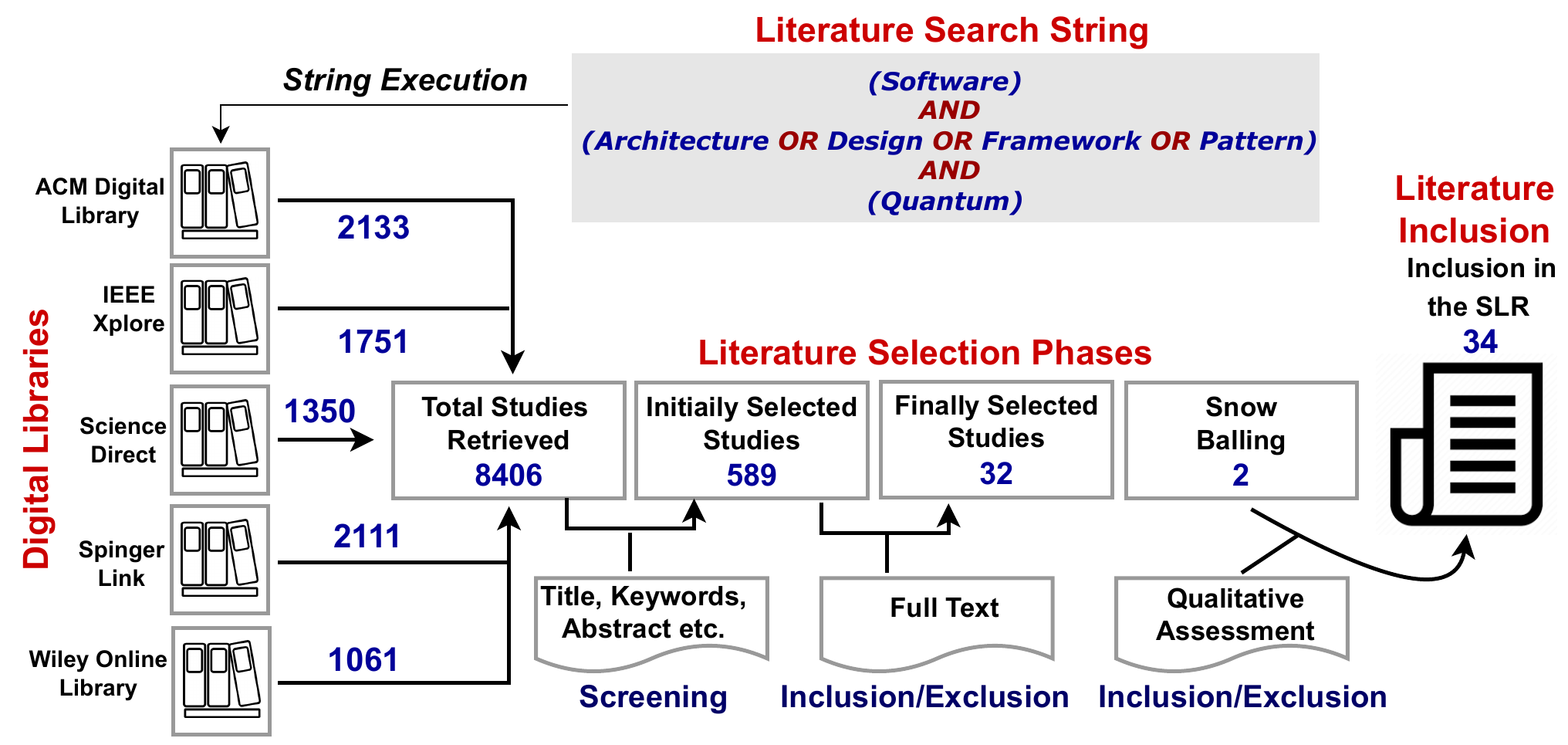}
 \caption{Studies selection process}
	\label{Fig:studiesselection}
\end{figure}

\subsubsection{Step-1: Select primary studies} \label{studies selection}
Primary studies search process started with exploring the selected digital repositories using the search string discussed in Section \ref{Search strategy}. The search process was initiated on 30th September 2021 and ended on 9th October 2021. Initially, the search string returned a total of 8,406 studies, which are further filtered by the first three authors based on the studies titles, keywords, and abstracts against the inclusion and exclusion criteria (see Figure \ref{Fig:studiesselection}). The second phase screening returned a total of 589 studies.
The third phase of inclusion and exclusion screening was performed based on the full-text review of the studies, where 32 primary studies were finally selected (see Figure \ref{Fig:studiesselection}). Additionally, the fourth and fifth authors were invited to confirm the search findings and list of selected studies. 

For example, we used the advanced search option for IEEE Xplore (\textsf{`Search Term'}) to execute the search string to identify published studies (in \textsf{`Full Text \& MetaData}). The search yielded a total of 32115 studies, majority of which focused on quantum systems in general and quantum hardware in particular. While trying to eliminate an exhaustive list of irrelevant studies, we interchanged the search parameter (from \textsf{`in Full Text \& MetaData'} to \textsf{`in Abstract}) and found 397 studies that missed some relevant studies that were discovered before the search parameter interchange. Therefore, we decided to manually scan through the 32115 studies after we applied further digital library-specific filtering to eliminate search results classified under \textsf{`Standards'}, \textsf{`Books'} and other alike categories to get a total of 1751 candidate studies from IEEE eXplore. Based on a similar approach, often digital library-specific filtering, we extracted and identified the candidate studies to proceed with their screening, inclusion/exclusion, and qualitative assessment, as in Figure \ref{Fig:studiesselection}.

\begin{table*}[!h]\footnotesize
 \centering
 \scriptsize
 \caption{Inclusion and Exclusion Criteria}
  \label{Tab:inclusionexclusion}
 \begin{tabular}{|p{0.8cm}|p{5.3cm}|p{0.8cm}|p{5.3cm}|}
 \hline
 \textbf{Code}&\textbf{Inclusion Criteria}&\textbf{Code}&\textbf{Exclusion Criteria}\\\hline
  
Inc1 & Studies that specifically focus on software architecture studies in quantum computing domain & Excl1 & Exclude grey literature and duplicate studies.\\\hline
    Inc2 & Peer-reviewed published research (e.g., conference proceedings, journal articles, workshop/symposium papers & Excl2 & If multiple studies are published in the same project, then consider the one with maximum contribution.\\\hline
    Inc3 & Peer-reviewed studies available in full-text.& Excl3 & Exclude studies that do not model or describe structure and/or behaviour of quantum software. \\\hline
    Inc4 & Reported in English language. &Excl4 & Exclude the studies that do not discuss any of the software architectural aspects as outlined in the RQs (e.g., process, patterns, notations, tools)\\\hline
\end{tabular}
\end{table*}

Moreover, the backward snowballing approach was used to manually search the references list of the selected 32 primary studies to identify additional studies that might have been missed during the search string-based review process \cite{wohlin2014guidelines}. The backward snowballing eventually returned two more studies that explicitly fulfilled the inclusion and exclusion criteria. The snowballing process was mainly performed by the first and second authors. Additionally, the third and fourth authors were invited to mutually verify the  findings reported by the first and second authors. To include studies that discuss software architecture, we specifically looked for architectural models (graphical notations, e.g., UML diagrams) or architectural specifications (descriptive notations, e.g., ADLs) that represent the structure or behavior of software system \cite{TSEADL}\cite{UML}. Finally, (32+2) studies are shortlisted (see Figure \ref{Fig:studiesselection}) to review, analyze and address the research questions based on their findings. The selected primary studies list is provided in Appendix A (Table \ref{tab:selectedstudies}). Furthermore, We included several non-peer-reviewed studies available on the arXiv open-access repository \cite{zhao2020quantum}\cite{nguyen2022software}\cite{IoTTSE}\cite{CSURBC}\cite{graef2021software}\cite{RQ1-11}\cite{perez2021modelling}\cite{dey2020qdlc}\cite{QuantumProspects}\cite{wang2022qusbt}\cite{junior2021systematic}\cite{khan2022agile} to complement the study’s overall findings. However, we did not include them in the primary studies list  (Appendix A – Table 11) as per the guidelines of SLR \cite{kitchenham2004evidence} approach. In addition to following the guidelines of the SLRs for literature inclusion \cite{kitchenham2004evidence}, our decision was also motivated by the fact that preprints are often subject to changes overtime, with several versions having the same title but differing content. Given the fast-paced research fields of quantum software engineering/architecture, preprints may contain errors or changes that could compromise the reliability of our SLR’s results and threaten its internal validity. Therefore, we excluded preprints from our list of selected primary studies to minimize the mentioned risk. For instance, one preprint changed \cite{abbott2018effects} its content four times within two years, highlighting the need for caution when incorporating preprints into a systematic review.

In addition to our adopted approach for automated search in electronic databases and backward snowballing to identify the relevant studies, several other approaches could be used. Some of these approaches include but are not limited to searching individual publication venues (e.g., conference proceedings, journal volumes), research group publications, and forward snowballing \cite{felizardo2016using}.  Specifically, forward snowballing  - searching for studies that cite the studies contained in the seed set - is found to be more useful in updating or extending an already conducted secondary study but is still prone to missing relevant literature. Jalali and Wohlin \cite{jalali2012systematic} investigates the application of snowballing approaches in SLRs and suggests that similarity in identified literature is expected to increase if both the backward and forward snowballing are performed since the overlap in the included papers would be greater. This influenced our decision to avoid forward snowballing, however; future extensions of this SLR can benefit from forward snowballing with an updated  seed list of studies \cite{felizardo2016using}.

\subsubsection{Step-2: Perform quality assessment (QAs)}\label{quality assessment}
The quality of the selected studies is evaluated based on the quality assessment criteria that aim to remove the research bias and evaluate the degree of significance and completeness of the selected studies \cite{keele2007guidelines}. The quality assessment guidelines provided by Kitchenham and Charters are followed to develop the assessment criteria (see Table \ref{Tab:Qualityassessment}) \cite{keele2007guidelines} . The criteria consist of five assessment questions, and each selected primary study assessed against these questions \textit{(QAs1-QAs5)}. Assigned score (1) if the primary study explicitly addressed the QAs questions and (0.5) points if the questions are partially addressed. Similarly, studies with no evidence of considering the assessment questions are given 0 point. The final quality assessment score for each primary study is the sum of the score assigned against each QAs question. The first author applied the assessment criteria and the results were further independently verified by second and third authors. We include those studies in the final list which had accumulative QAs score greater than or equal to 1.5 \cite{waseem2020systematic}. The accumulative final score of each primary study against the QAs questions is given in Appendix A (Table \ref{tab:selectedstudies}).

\begin{table*}[!h]\footnotesize
 \centering
 \scriptsize
  \caption{Studies quality assessment criteria}
  \label{Tab:Qualityassessment}
  \begin{tabular}{|c|p{11cm}|l|p{3cm}|}
  \hline
\textbf{Code} & \textbf{Quality Assessment Questions} & \textbf{Score} \\\hline

    QAs1	& Do the research objectives of the study are explicitly defined? & (1/0.5/0) \\\hline
    QAs2	& Does the adopted research methodology is clearly discussed?	& (1/0.5/0) \\\hline
    QAs3	& Do the experimental settings are explicitly reported? & (1/0.5/0) \\\hline
    QAs4	& Do the results and findings are thoroughly discussed? & 	(1/0.5/0) \\\hline
    QAs5	& Do the real-world implications of the study are reported? &	(1/0.5/0) \\\hline
  \end{tabular}
\end{table*}

\subsubsection{Step-3: Perform data extraction} \label{perform data extracion}
We defined a set of data extraction items (see Table \ref{Tab:Dataitems}) to address the RQs formulated in Section \ref{RQs}. Data items are the particular types of data extracted from each selected primary study that directly map to the study RQs. The first author performed the pilot data extraction process for ten studies to evaluate the reliability of the extracted data items. The second and third author assessed the pilot study findings, and based on their suggestions, the first author revised the data extraction items. The formal data extraction process was performed by the first three authors by equally distributing the total number of selected primary studies, and the studies distribution was done based on the authors' research expertise and interest. The general (demographic) details of each selected primary study were extracted against the data items (DI1-DI4), and the rest (DI5-DI13) are specific to the study RQs.

We finally conducted the Cohen's Kappa test to check inter-personal bias in the primary studies selection (Section \ref{studies selection}), quality assessment (Section \ref{quality assessment}), and data extraction (Section-\ref{perform data extracion}) phases. Mainly, the first three authors were involved in the studies selection, quality assessment, and data extraction process. To remove the inter-personal bias for the mentioned phases of the SLR process, we invited the remaining authors and merged them across two different groups (authors 4-5, authors 6-7). They were asked to randomly select a set of ten primary studies and sequentially perform the studies selection, quality assessment, and data extraction process as performed by authors 1-3.  Eventually, the Cohen's Kappa test was performed to measure the agreement level and identify the significant differences across the mentioned phases between all the three groups of authors (authors 1-3, authors 4-5, authors 6-7). The Cohen’s Kappa test is widely adopted in EBSE research \cite{perez2020systematic}. Cohen's kappa coefficient (k) is \textit{the proportion of chance-expected disagreements which do not occur, or alternatively, it is the proportion of agreement after chance the agreement is removed from consideration} \cite{cohen1960coefficient}. The (k) coefficient measures the level of agreement between a group of raters that evaluate N-objects into (c) mutually exclusive categories \cite{cohen1960coefficient}. The agreement level between the raters equals chance agreement when Cohen's kappa coefficient value (k)=0. The level of agreement is positive when (k) is greater than the chance agreement and negative if it is less than it. The perfect agreement occurs between a group of raters when the value of k ranges from (0.81 to +1.00). The interpretation of the k-value to measure the strength of agreement is adopted from the observer agreement study conducted by Landis and Koch \cite{landis1977measurement} 

We used R-3.6.3 to conduct the (k) test for interpreting the agreement level between the groups of raters (authors). The R-code (see Appendix A (Table \ref{tab:Cohens Kappa test R-Code})) was executed and obtained the Cohen's Kappa coefficient value (k= 0.62), which shows a positive and substantial agreement level \cite{landis1977measurement} between all the authors for the primary studies selection, quality assessment, and data extraction phases.  Based on the test results, we concluded that no personal bias exists between the authors that could significantly impact the core SLR phases.\\

Similarly, the Cohen’s Kappa test was performed to evaluate the interpersonal bias between the authors for the snowballing process. Mainly authors 1-4 performed the snowballing process (Section \ref{studies selection}), however, authors 5-6 were  invited to participate in the Cohen’s Kappa test to assess inter-personal bias. Both groups (authors 1-4, authors 5-6) selected the first five primary studies (S1 to S5) and performed the snowballing search. The Cohen’s kappa coefficient (k= 0.50) is calculated based on the search findings of both groups using the R-code given in Appendix A (Table \ref{tab:Cohens Kappa test R-Code}). The given value of (k) reveals that both groups of authors have an unbiased, positive, and moderate level of agreement for the snowballing process.\\

\begin{table*}\footnotesize
 \centering
 \scriptsize
  \caption{Relevant data items extracted from the selected primary studies}
  \label{Tab:Dataitems}
  \begin{tabular}{|p{0.8cm}|p{3cm}|p{6cm}|p{2cm}|}
        \hline
\textbf{Code} & \textbf{Data item} & \textbf{Description} & \textbf{Related RQ} \\
\hline
    QI1	& Index & The study ID & Demographic \\\hline
    QI2	& Study title & Full title of primary study & Demographic \\\hline
    QI3	& List of authors & Authors full names &  Demographic\\\hline
    QI4	& Publication’s venue & Name of the Journal, Conference, Workshop, Book, symposium, Magazine &  Demographic\\\hline
    QI5	& Publication’s year & Temporal information of each study. &  RQ 1.1\\\hline
    QI6	& Publication type & Journal, Conference, Workshop, Book chapter, Magazine &  RQ1.1\\\hline
    QI7	& Research type & Studies mapping across research facets &  RQ1.2\\\hline
    QI8	& Research domain & Develop themes and sub-themes of studies research focus across different domains &  RQ1.3\\\hline
    QI9	& Architectural activities & Key activities to define quantum software architecture process &RQ2.1\\\hline
    QI10 & QSA modeling notations & The existing modeling  notations to structure quantum software architecture & RQ2.2\\\hline
    QI11 & QSA patterns &  Identify the patterns for quantum software architectural design problems & RQ2.3\\\hline
    QI12 & Architectural tools and frameworks & The tools discussed in the primary studies to support architecting activities & RQ2.4\\\hline
    QI13 & QSA challenges & The challenges reported to develop quantum software and system architecture & RQ2.5 \\\hline
  \end{tabular}
\end{table*}

\subsubsection{Step-4: Perform data synthesis} \label{Synthesis}
Data items (DI1-DI4) were analysed using the descriptive statistical approach. Similarly, we generated initial codes for the data items (DI7, DI8, DI12 and DI13) to define the research themes and address RQ1.2, RQ1.3,  RQ2.4 and RQ2.5. Thematic analysis guidelines for qualitative data provided by Braun and Clarke are considered to systematically analyze, organize, and develop  themes across the extracted data \cite{braun2006using}. In line with the outlined RQs (\ref{Tab:RQsTable}), the following thematic data analysis steps are followed to develop the key themes of extracted data items:

\begin{enumerate}
 \item \textbf{Data familiarization}: The first three authors thoroughly read the selected primary studies and noted the data items given in Table \ref{Tab:Dataitems}.
    \item \textbf{Generating the initial codes}: The initial codes from the extracted data are generated to define the research themes for RQ1.2, RQ1.3, RQ2.4, and RQ2.5.
    \item \textbf{Searching for themes}: The codes define in the previous step are analyzed and encapsulated across broader themes.
    \item \textbf{Reviewing themes}: The first three authors examined the themes to separate, drop and merge based on the mutual discussion and understanding.
    \item \textbf{Defining and naming themes}: The defined themes are characterized with precise names.
    \item \textbf{Producing the report}: This step involves to refine the developed themes and their respective characteristics.
\end{enumerate}

\begin{figure}[!htbp]
 \centering
 \includegraphics[scale=0.60]{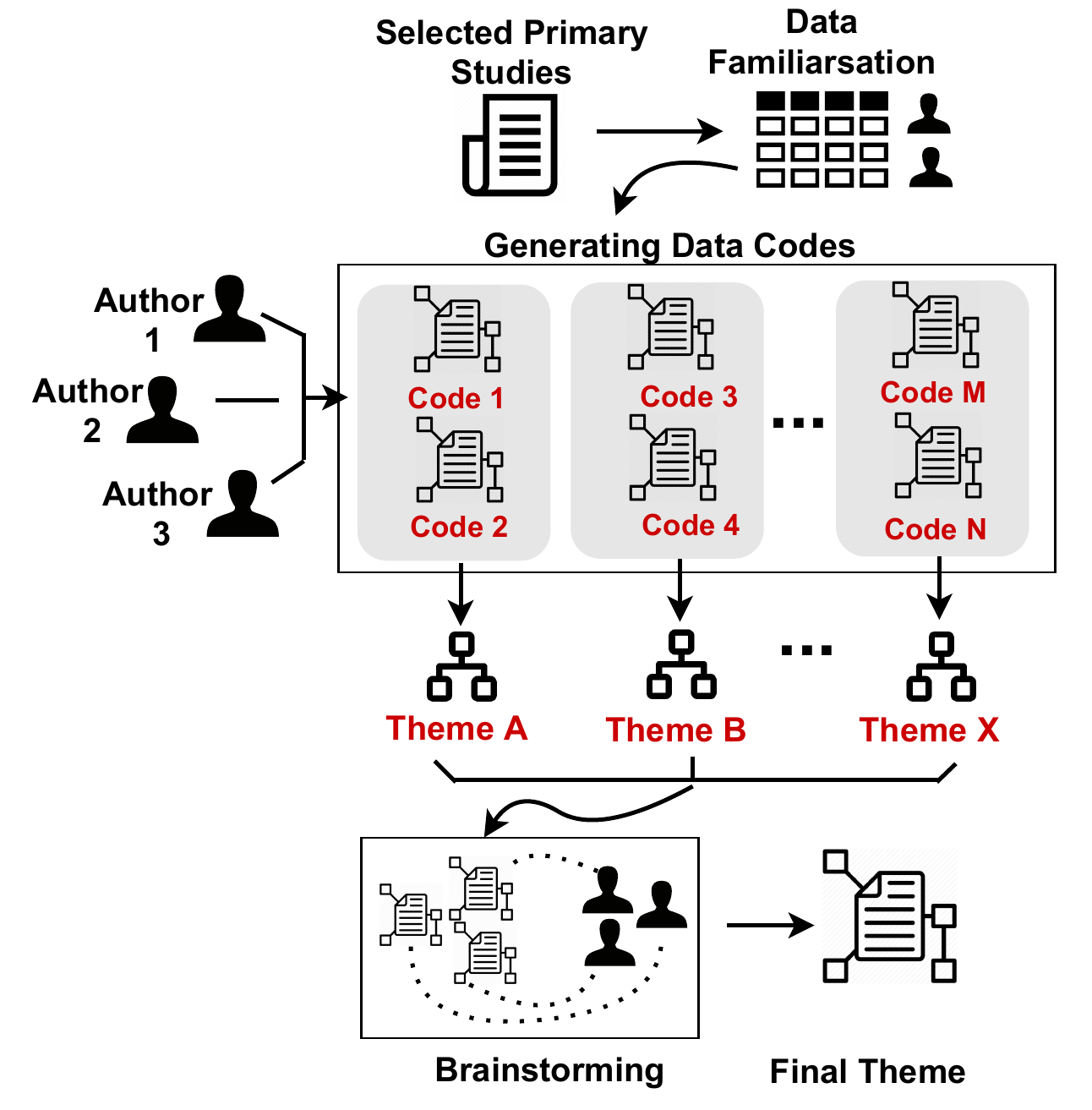}
 	\caption{Thematic analysis process}
	\label{Fig:thematicanalysis}
\end{figure}

The thematic analysis process of this SLR is given in Figure \ref{Fig:thematicanalysis}, and all the authors finally participated in the brainstorming session to remove bias in the thematic approach by defining and naming the key themes. To complement the methodological steps of this SLR, a replication package is provided  that details the selected primary studies based on the customised search string, scoring of quality assessment of identified studies, and the extracted data for each individual RQs \cite{ReplPackage}.

\subsection{Reporting the review}
 We reported the results of SLR, presented in dedicated sections, based on the categories of outlined RQs (see Table \ref{Tab:RQsTable}). Specifically, (i) \textit{demography details} of published research (i.e., RQ1.1 to RQ1.3) are discussed in Section \ref{demographyresults}, and (ii) \textit{architectural solutions and challenges}  (i.e., RQ2.1 to RQ2.5) are detailed in Section \ref{RQ2}. The analysis of the SLR and summary of key results are detailed in Section \ref{keyfindings}. 
\section{Demography Details of Published Research} \label{demographyresults}
In this section, we answer RQ1, having three sub-questions, i.e., RQ1.1 - RQ1.3 that rely on mapping analysis to present demography details of published research \cite{SMS}. Specifically, within the SLR, we performed systematic mapping to investigate demography details of published research focus on (i) types and frequency of publications (RQ1.1: Section 4.1), (ii) types and contributions of research studies (RQ1.2: Section 4.2), and (iii) application domains of architectural solutions (RQ1.3: Section 4.3), all detailed below. The demography details complement the presentation of overall results and discussion of proposed architectural solutions. For example, the types of research studies (answering RQ1.2) discussed here indicate a multitude of research contributions, such as solution proposals, validation research, and/or philosophical studies, and their roles in deriving the architecting activities for quantum software.      

\subsection{Types and frequency of publications (RQ1.1)} \label{RQ1.1}
\begin{figure}[!t]
 \centering
 \includegraphics[scale=0.75]{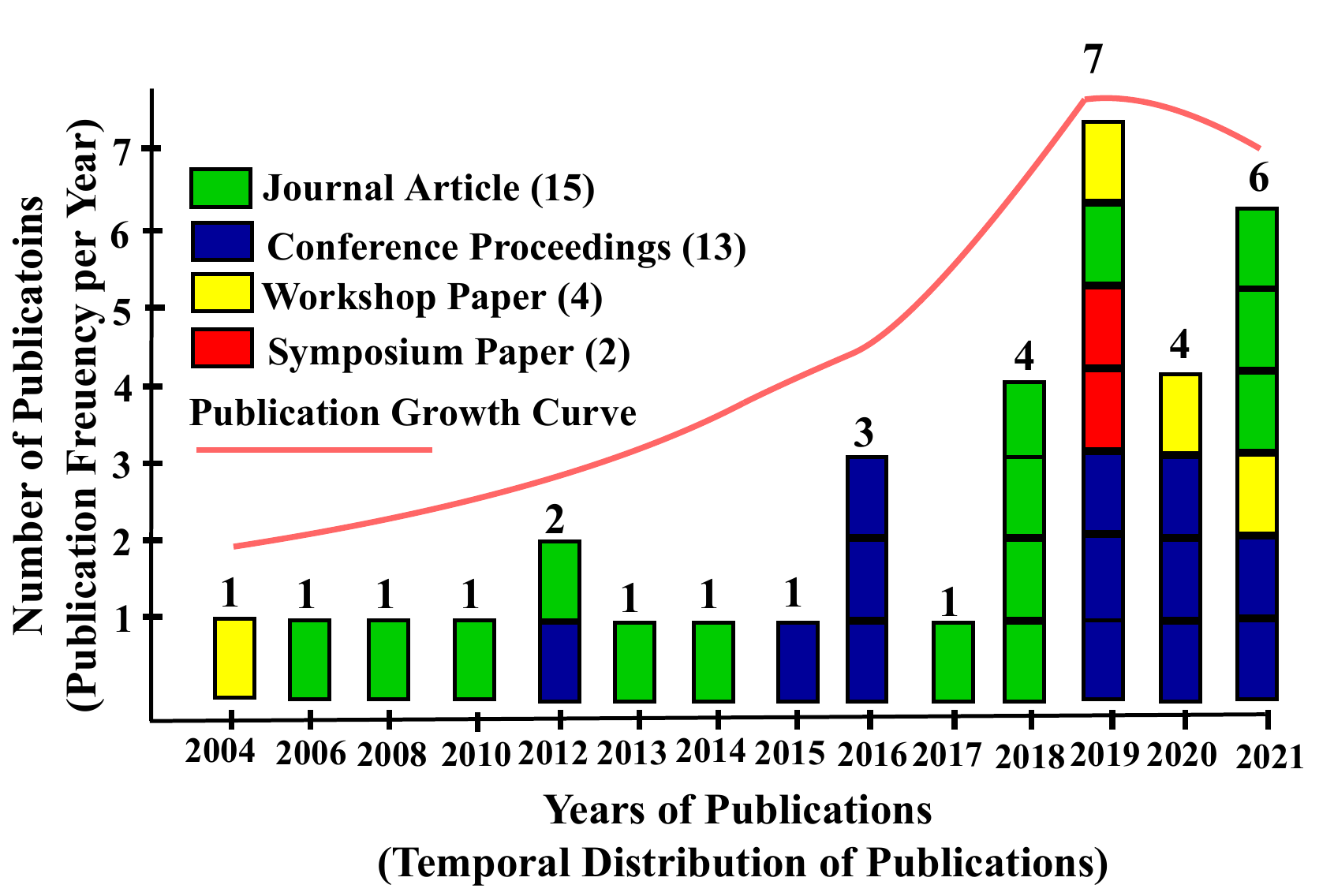}
 	\caption{Overview of Frequency and Types of Publications}
	\label{Fig:Frequencyandpublicationtype}
\end{figure}

It is significant to classify the selected primary studies based on their frequency and type of publications. This analysis highlights the research trend of a particular research area and the research community's interest. The frequency indicates how frequent is the occurrence of publications over the years, whereas the types refer to a specific type of publications (e.g., a journal article) as illustrated in Figure \ref{Fig:Frequencyandpublicationtype}. The total number of published studies are presented across (Y-axis) and their year of publication across (X-axis). Moreover, Figure \ref{Fig:Frequencyandpublicationtype} is a bar graph that relatively highlights different publication types, i.e., conference proceedings, journal articles, symposium papers, and workshop articles. The initial study was published in 2004 and final in October 2021. The bar graph reveals that a total 21 (62\%) of the selected studies were published in the last four years (from 2018 to October 2021), which is an interesting finding that interprets the significance of quantum software architecture in present-day quantum computing research. It reveals that the research community are significantly working on designing architectural solutions for quantum software systems. Moreover, 15 (44\%) primary studies are published in journals, 13 (38\%) in conference proceedings, 4 (12\%) workshop papers and 2 (6\%) symposium article. A report by Scopus in 2021 highlights recent research trends on quantum computing, reflected via Scopus-indexed documents, in terms of demography details of published research. The report provides a multi-faceted overview of published research regarding frequency, types, top institutes, and top contributors regarding their research on quantum computing \cite{Quantum-Scopus}.

\begin{tcolorbox}[colback=gray!5!white,colframe=gray!75!black,title=Key Findings of RQ1.1]
\textbf{Finding 1}: Maximum number of primary studies (n=21, 62\%) are published from 2018 to 2021. It exhibits that quantum software architecture is emerging research area and got significant attention of research community.

\textbf{Finding 2}: Regarding publications type, the given results underline that journals (n=15, 44\%)  and conferences (n=13, 38\%) are the popular venues to publish the relevant studies.
\end{tcolorbox}

\subsection{Types of research and contributions (RQ1.2)} \label{RQ1.2}
The selected publications are categorised based on the following six well-established research types proposed by Wieringa et al.~\cite{wieringa2006requirements}: \textit{evaluation research}, \textit{proposal of solution}, \textit{validation research}, \textit{philosophical papers}, \textit{opinion papers}, and \textit{personal experience papers}. \textit{Evaluation research} is conducted to evaluate a specific problem or solution in practice using different empirical research techniques. \textit{Proposal of solution} articles develop a method or solution for a relevant problem without fully validating its significance. \textit{Validation research} is conducted to evaluate the quality attributes of the proposed solution, which has not yet been deployed in a real-world environment. \textit{Philosophical papers} focus on architecting theoretical or conceptual frameworks. \textit{Opinion papers} discussed authors' negative or positive opinions regarding a specific framework, model, a solution. In \textit{Personal experience papers}, the authors report their personal experiences regarding a particular project or group of it. Additionally, we reported the research contribution of each paper classified across the mentioned research types.

\begin{figure}[!t]
 \centering
  \includegraphics[width=\linewidth]{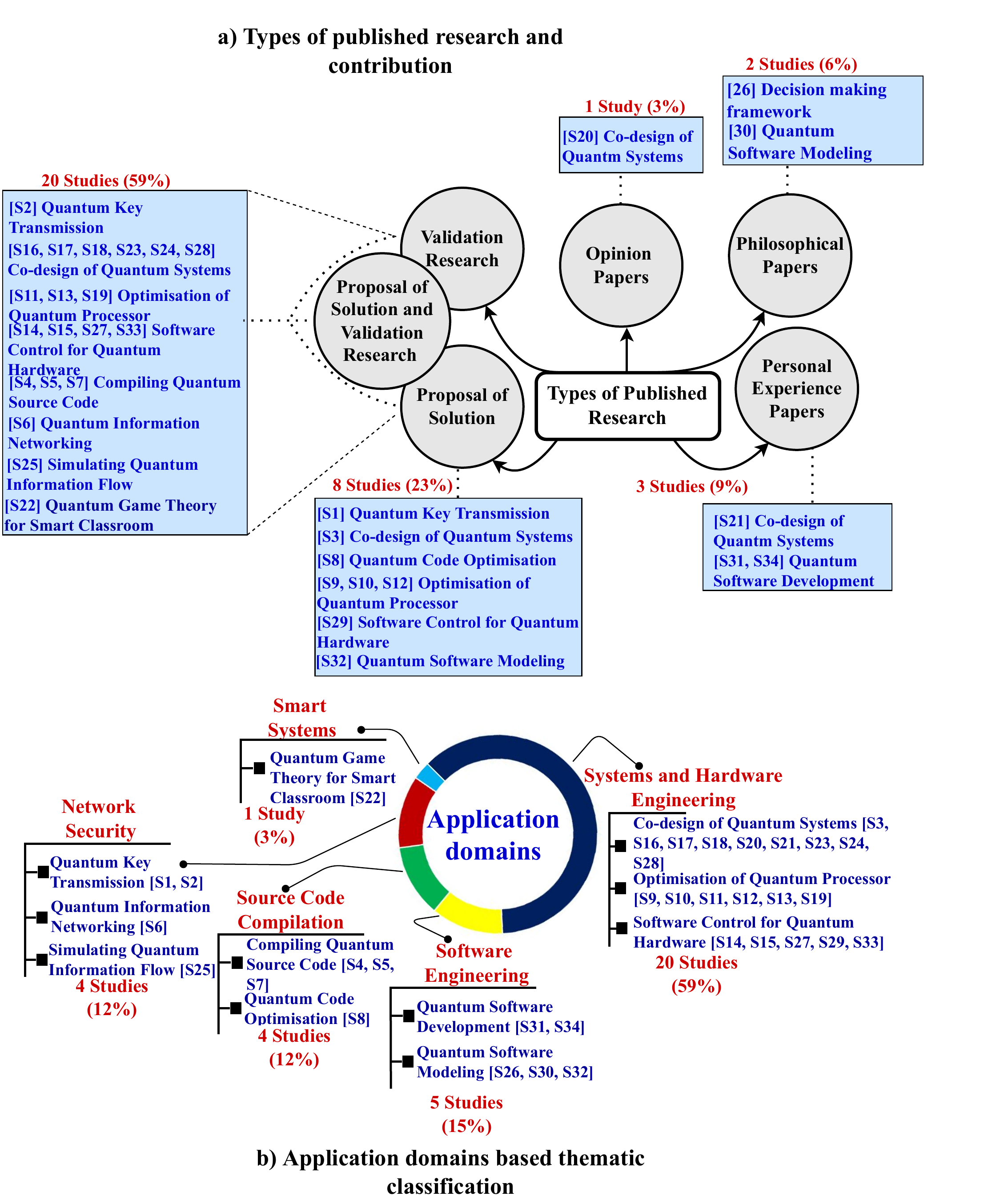}
 	\caption{Overview of Types of Research and Application Domains}
	\label{Fig:Researchfacetsandfocus}
\end{figure}

Thematic analysis process discussed in Section \ref{Synthesis} is followed to address RQ1.2 and classify the selected 34 primary studies across the given research types (see Figure \ref{Fig:Researchfacetsandfocus} (a)). The set of selected studies consist of \textit{(n=8, 24\%) proposal of solution}, \textit{(n=3, 9\%) personal experience papers}, \textit{(n=2, 6\%) philosophical papers}  and \textit{(n=1, 3\%) opinion papers}. Moreover, we identified (n=20, 59\%) studies that cover both proposal of solution and validation research categories. These studies are classified in a separate category (i.e., \textit{proposal of solution and validation research}) (see Figure \ref{Fig:Researchfacetsandfocus} (a)). We did not identify any paper that fits in the evaluation research category; therefore, it is excluded from the mapping process.

The results given in Figure \ref{Fig:Researchfacetsandfocus} (a) reveal that majority of the studies, i.e., \textit{(n=20, 59\%)} are mapped in the heterogeneous (i.e., \textit{proposal of solution and validation research}) category. It means that the selected studies proposed their own solutions and conducted sample implementation to validate the significance of those solutions. It is aligned with the fact that quantum software architecture is a new paradigm, and there is a demanding need of novel architecture solutions.
The second most common category is \textit{proposal of solution}  \textit{(n=8, 23\%)}, where various architectural solutions are proposed. However, the proposed solutions are not evaluated or validated both empirically and in real-world practice. For example, software architecture is proposed in [S1]\footnote{Please note, the notation [$S_{n}$], where n represents a numerical value (range: 1 to 34) to indicate a reference to the selected primary studies for SLR, listed in \textbf{Appendix A (Table \ref{tab:selectedstudies})}. This notation also help to distinguish the selected primary studies from references in the bibliograpay section of this paper.} to set up an ecosystem for quantum key distribution (QKD) in quantum networks. The architecture is build using a set of modules i.e., QKD module, relay modules, and QKD node. However, the proposed solution is not validated or evaluated to assess its real-world implications and contributions.
Three \textit{(n=3, 9\%)} primary studies are mapped into the \textit{personal experience papers} category [S21, S31, S34]. For example, Leymann et al. [S31] reported an understanding of the architectural model to support business processes for developing and sharing the quantum software systems. 
 The \textit{philosophical papers} category covers two \textit{(n=2, 6\%)} primary studies [S26, S30]. For instance, Nallamothula proposed a theoretical decision-making framework for quantum software architecture selection [S26]. The proposed framework has not been evaluated experimentally or in real-world practice. One single study \textit{(n=1, 3\%)} is categorised as \textit{opinion paper}, where the authors shared the opinion of quantum and classical co-design architecture [S20]. 
Regarding overall contribution, we noticed that 9 studies focused on quantum-classical intersection (i.e., \textit{co-design of quantum systems}), where both classical and quantum techniques used to develop the quantum software architecture (see Figure \ref{Fig:Researchfacetsandfocus} (a)) [S3, S20, S21, S16, S17, S18, S23, S24, S28]. It is a known fact that quantum software development is not a well establish field. Presently, its not possible  to entirely develop a quantum software architecture based on quantum computing concepts. We still need to consider the classical software development concepts and techniques, at least at the interface level, to structure a quantum software system.

\begin{tcolorbox}[colback=gray!5!white,colframe=gray!75!black,title=Key Findings of RQ1.2]
\textbf{Finding 3}: Analysing the types of published research highlight that the combination of solution proposals and validation research represent most frequent publications. A total of 20 studies (i.e., n=20, 59\% approx) represent this category to propose architectural solutions and validate quantum software solutions via simulation or case studies. 

\textbf{Finding 4}: Thematic classification of the research contributions highlights that (n=9, 26\% approx) studies focused on proposing architectural solutions for \textit{co-design of quantum systems}. Quantum system co-design refers to mapping between classical and quantum concepts during the development of quantum computing systems. It means that most of the studies focused on developing quantum software systems using both classical and quantum computing concepts. 
\end{tcolorbox} 

\subsection{Classification of application domains (RQ1.3)} \label{thematic}
Thematic process defined in Section \ref{Synthesis} is followed to categorise the selected primary studies based on the common application domains. Systematic identification, categorization, and naming process of identified themes and sub-themes are given in Figure \ref{Fig:Researchfacetsandfocus} (b). 

We collected at least two or more studies of common application domains and encapsulate them under a single umbrella called theme. In this study, the following five core themes are identified and the selected studies are classified across them: (i) \textit{systems and hardware engineering (n=20, 59\%)}, (ii) \textit{software engineering (n=5, 15\%)}, (iii) \textit{smart systems (n=1, 3\%)}, (iv) \textit{source code compilation (n=4, 12\%)}, and (v) \textit{network security (n=4, 12\%)}.
In sub-thematic classification, we further categorised the main themes into more specific topics. Sub-themes are secondary to core themes, where the overall application domain (core theme) is classified more narrow (sub-themes). For example, the core theme (\textit{systems and hardware engineering}) is classified across three distinct sub-themes including \textit{co-design of quantum systems (n=9, 26\%)}, \textit{optimisation of quantum processor (n=6, 18\%)}, and \textit{software control for quantum hardware (n=5, 15\%)}. Similarly, \textit{software engineering} is sub-classified into \textit{quantum software development (n=2, 6\%)} and \textit{quantum software modeling (n=3, 9\%)}. \textit{Source code compilation} has two sub-themes: \textit{compiling quantum source code (n=3, 9\%)} and \textit{quantum code optimisation (n=1, 3\%)}. Moreover, the classification given in Figure \ref{Fig:Researchfacetsandfocus} (b) illustrates that \textit{network security} is further categorised across sub-themes \textit{quantum key transmission (n=2, 6\%)}, \textit{quantum information networking (n=1, 3\%)}, and \textit{simulating quantum information flow (n=1, 3\%)}. Finally, the core theme \textit{smart systems} has only one sub-theme i.e., \textit{quantum game theory for smart classroom (n=1, 3\%)}.

Figure \ref{Fig:Researchfacetsandfocus} (b) provides the high level categorisation of the existing quantum software architectural solutions with respect to different application domains. For instance, \textit{systems and hardware engineering} is the most common and explicitly explored application domain with twenty research studies. It is aligned with the fact that the research focus on quantum software development is heating up \cite{zhao2020quantum}. Technology giants e.g., Google, Alibaba, and IBM are marching forward to propose advance architectural solutions to take the lead in quantum software technologies \cite{WinNTMark}. Similarly, \textit{co-design of quantum systems} is a sub-theme of \textit{systems and hardware engineering}, which has a total of nine studies. It highlights the significance of quantum-classical hybridization. Quantum-classical collaborative relationship will have significant impact on quantum software architecture in the near-term \cite{baczewski2017co}. It will improve the architectural efficiency and meet the require performance.

In summary, Figure \ref{Fig:Researchfacetsandfocus} (b) provides a holistic overview of studies mapping with respect to the application domains. It enables different interpretations of published studies based on core research themes and sub-themes. The given mapping provides a taxonomical understanding of state of the art application domains.

\begin{tcolorbox}[colback=gray!5!white,colframe=gray!75!black,title=Key Findings of RQ1.3]
\textbf{Finding 5}: The core application domains are: \textit{systems and hardware engineering , software engineering , smart systems, source code compilation,} and \textit{network security}. The selected primary studies are categorised across the mentioned domains.

\textbf{Finding 6}: \textit{Systems and hardware engineering } (n=20, 59\%) is identified as the most common application domain. It reveals the fact that research community significantly focuses on presenting architectural solutions for quantum system and hardware problems.  The reason might be that the existing classical system engineering approaches are not able to explicitly encompass the attributes of quantum physics \cite{everitt2016quantum}. There is a need of novel system and hardware engineering frameworks that tackle the quantum interface problems.

\end{tcolorbox}

\section{Architecture-Centric Solutions for Quantum Software} \label{RQ2}

We now discuss architecture-centric solutions and emerging challenges, answering RQ2.1 to RQ2.5, that highlight some of the core aspects of designing and implementing quantum software. Specifically, (i) we present architectural process and its underlying activities (RQ2.1: Section \ref{RQ2.1}), (ii) architectural modeling notations (RQ2.2: Section \ref{RQ2.2}), (iii) architectural patterns and design decisions (RQ2.3: Section \ref{RQ2.3}), (iv) tools and frameworks (RQ2.4: Section \ref{RQ2.4}), and challenges of quantum software architecture (RQ2.5: Section \ref{RQ2.5}). 

\subsection{Architectural process and activities (RQ2.1)} \label{RQ2.1}
\begin{landscape}

\begin{figure}[h!]
 \includegraphics[scale=0.58]{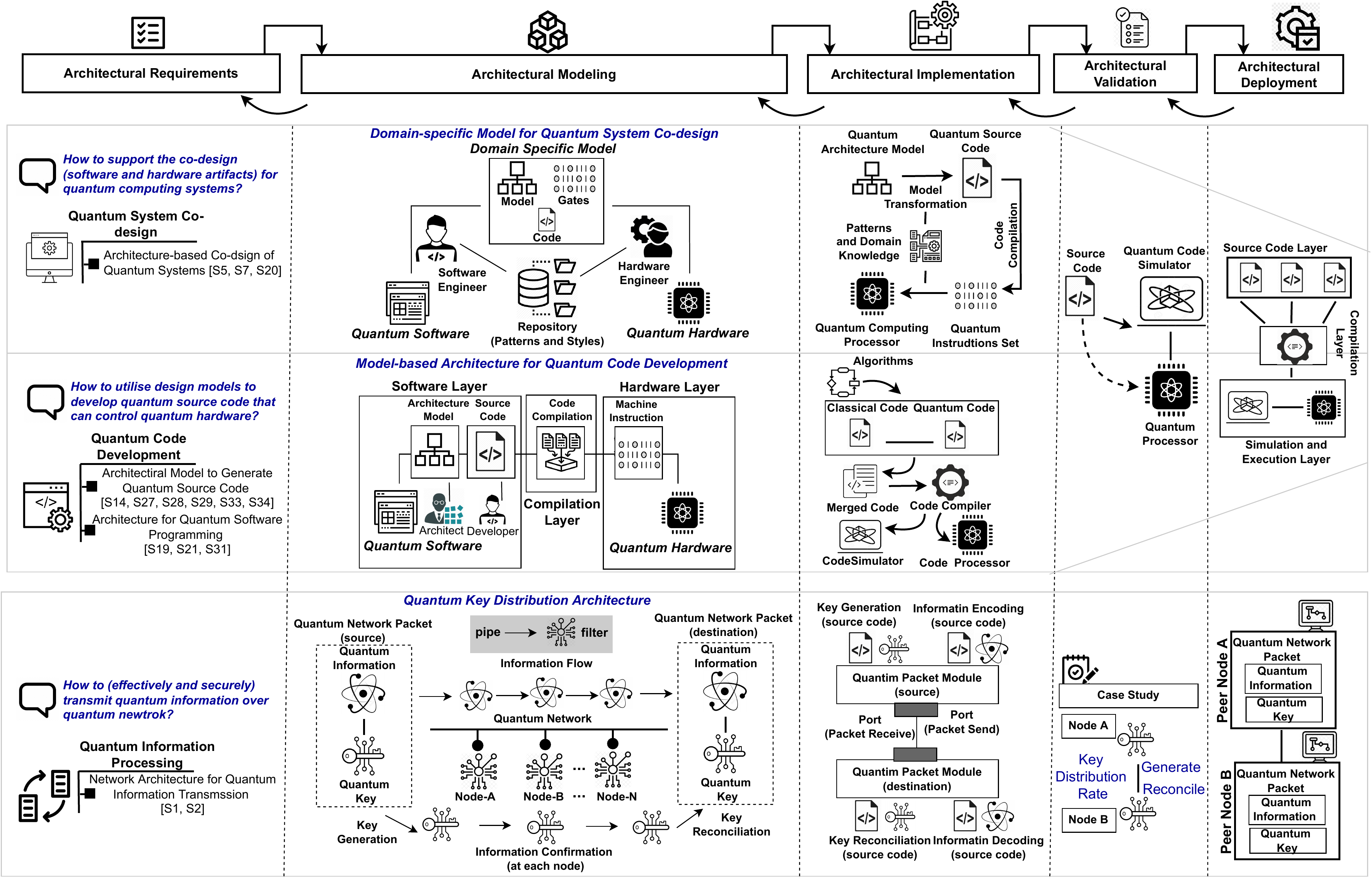}
 	\caption{Overview of Quantum Architecting Process and its Activities}
	\label{QSA}
\end{figure}
\end{landscape}

We now answer RQ2.1 that aims to investigate the existing process(es) that can support a process-centered - incremental and structured - approach to architect quantum software systems \cite{RQ1-1}. Specifically, an architectural process comprises of a collection of activities (a.k.a. architecting activities to support analysis, synthesis, and evaluation of the architecture for quantum software systems and applications \cite{RQ1-2}\cite{svore2006layered}. During system design and implementation phase, architectural process streamlines \textit{what} needs to be done and provides an umbrella to accumulate a collection of architecting activities that demonstrate \textit{how} it is to be done. For example, in an architectural process, the activity called architectural requirements aims to analyse and outline the design challenges/issues that a particular architecture must resolve. The outcome of architectural analysis activity is a set of architecturally significant requirements (ASRs) to highlight the needed functionality and desired quality of the system under design \cite{RQ1-2}. For example, as in Figure \ref{QSA}, as part of architectural requirements one of the ASR is: \textit{how to effectively and securely transmit quantum information over quantum network?} The ASR outlines a design challenge that must be addressed by designing the appropriate architecture that supports transmission of quantum information (i.e., required functionality) over quantum network in an efficient and secure manner (i.e., desired quality attribute). The relevant studies, as an evidence, that support architectural process are indicated in Figure \ref{QSA}. For example, Figure \ref{QSA} shows that two studies specify the requirements of an reference architecture, as a software blueprint, to generate quantum source code [S14, S27].  

From quantum software engineering perspective, existing architectural processes represent a concentrated knowledge and wisdom (derived from architects’ experiences, industrial practices, and academic solutions that can be attuned to architectural challenges for quantum genre of software systems) \cite{zhao2020quantum}\cite{RQ1-1}\cite{RQ1-5}\cite{RQ1-6}. However, architecting quantum systems entail some specific challenges that cannot be effectively addressed by existing processes that have been designed for classical computing systems. Some of the quantum specific challenges include but are not limited to co-design, i.e., mapping quantum algorithms to Qubits of a Qugates, compiling hybrid source code into a unified quantum instruction set, and configuring simulators to simulate and execute quantum code \cite{svore2006layered}\cite{RQ1-7}. This means that existing architectural processes need customised activities to address design challenges of quantum software. 

To present the results, we followed available guidelines and empirically-based studies, grounded in industrial practices and academic research to document software architectures in terms of architectural processes and their underlying architecting activities \cite{RQ1-1}\cite{RQ1-2}\cite{RQ1-5}\cite{RQ1-6}. We followed a generic process pattern derived from five industrial approaches to document architectural processes in terms of architectural design activities namely architectural analysis, architectural synthesis, and architectural evaluation \cite{RQ1-1}. The architectural process model proposed by Hofmeister et al. \cite{RQ1-1} is incorporated by Tang et al. \cite{RQ1-8} with two additional activities namely architectural implementation and architectural maintenance. Some industrial surveys, incorporating practitioners’ perspective also highlight the needs for fine-grained representation, specifically in the context of architectural synthesis activity to effectively represent architectural solutions \cite{RQ1-5}. To support a fine-granular representation of the architectural synthesis activity, we divided it into two distinct activities namely architectural modeling (representing ASRs as an architectural model) and architectural implementation (transform architectural model into specifications that can be executed or simulated). In the following, we detail the architectural process for quantum software, defined in terms of architecting activities, illustrated in Figure \ref{QSA} that also acts as a running example for demonstrative purposes. Figure \ref{QSA} provides a visual catalogue of the process and activities that are exemplified based on the available evidence from the reviewed literature. During the review, each study that corresponds to an architecting activity was identified, whereas Figure \ref{QSA} was constructed by synthesizing the overall contributions from a collection of studies for their generic representation as a unified architectural process.

For example, as per Figure \ref{QSA}, the selected studies help us to identify the \textbf{architectural requirements} to support efficient and secure transmission of quantum information over a quantum network [S1, S2]. The proposed \textbf{architectural model} as in Figure \ref{QSA} relies on a pipe and filter architectural pattern that supports generation, transmission, and reconciliation of a quantum key to secure quantum information that travels over quantum network \cite{RQ1-10}. To support the modeling, \textbf{architectural implementation} is enabled via components, representing computational units for source (transmitter) and target (receiver) nodes in the network that coordinate quantum information via component ports. A case study based approach is adopted for architectural validation in terms of efficiency and security of generating, transmitting, and reconciling the quantum key [S1, S2]. Peer to peer configuration of network nodes is adopted for \textbf{architectural deployment}.

\begin{enumerate}[label=(\roman*)]
 \item Architectural Requirements as the initial activity of the process aims at analysing, filtering, and/or reformulating architectural concerns to derive a set of architecturally significant requirements (a.k.a., architectural requirements). This activity aims to define the problems that an architecture needs to address.
    \item Architectural Modeling aims to satisfy the identified architectural requirements by creating an overall architecture of the system that acts as a blue-print for the implementation. This activity represents the first steps towards providing an architectural solution for ASRs, while bridging the gap between requirements (i.e., desired functionality and quality) and implementation (i.e., executable or simulatable specifications).
    \item Architectural Implementation exploits the architectural model to implement the software system in terms of algorithmic specifications and executable source code. The implemented software relies on programming languages, compilers, and tools to write, compile, and execute the software.  
    \item Architectural Validation focuses on validating the functionality and quality of the implemented software in the context of architectural requirements. Architectural validation assesses the extent to which the required functionality (i.e., functional requirements) and desired quality (i.e., non-functional requirements) are being satisfied by the implemented software. 
    \item Architectural Deployment as the last activity of life cycle is concerned with deploying the validated software for its operationalisation. The deployment involves configuring the executable specification (architectural implementation) on a deployment node (typically an application server) that facilitates the execution of the deployed software.
\end{enumerate}

Based on the available evidence, as illustrated in Figure \ref{QSA}, the architectural process and its underlying architecting activities for classical computing systems can be tailored to support the architecting process for quantum software systems. However, architectural modeling and implementation activities must explicitly cover architectural requirements specific to quantum software \cite{RQ1-1}\cite{RQ1-2}\cite{RQ1-8}. For example, the \textbf{architectural requirement} in Figure \ref{QSA}, i.e., quantum system co-design requires analyzing and selecting the hardware (e.g. quantum processor) as well as software (e.g., quantum search algorithm) components to effectively design a quantum computing system (Figure \ref{QC}) [S5]. To satisfy this requirement, software as well as hardware engineer need a collaborative design of \textbf{architectural model}, referred to as a domain specific model that incorporates software architectural components mapped to instruction set for quantum computing processor [S7]. The co-designed model for a quantum computing system requires architectural \textbf{implementation} via model transformation. Model transformation exploits the concepts of model-driven architectures to transform architectural model into the high-level source code that is compiled into quantum instruction set by means of model traceability (mapping between architectural model and executable instruction set) and mode transformation (transition from architectural model to executable instruction set) [S20]. As in Figure \ref{QSA}, the architecting activities can be iterative, for example, in case of any mismatch between architectural model (i.e., design) and instruction set (i.e., execution) at architectural implementation phase requires maintenance or refactoring of the domain specific model at architectural modeling phase to ensure consistency between design and implementation.

\begin{tcolorbox}[colback=gray!5!white,colframe=gray!75!black,title=Key Findings of RQ2.1]
\textbf{Finding 7}: An architecture design endeavour for the quantum software requires an architecting process to incorporate a number of architecting activities. Existing architectural process can be leveraged to support five architecting activities for quantum software namely \textit{(i) architectural requirements, (ii) architectural modeling, (iii)  architectural implementation, (iv) architectural validation,} and \textit{(v) architectural deployment}.

\textbf{Finding 8}: Quantum specific requirements such as modeling Qubits to Qugates and co-design of quantum hardware and software requires domain specific modeling and transformation to be supported by architectural process activities.
\end{tcolorbox}

\subsection{Architectural modeling notations (RQ2.2)} \label{RQ2.2}
We now answer RQ2.2 that investigates the modeling notations, representing a multitude of graphical models or descriptive notations to specify, document, or represent the architectural models. From architectural process perspective (RQ2.1), the terms modeling notation, modeling language, and architectural language are virtually synonymous and often used interchangeably all referring to same concept of architectural representation either graphically or textually \cite{RQ1-5}\cite{perez2021modelling}. For example, to support quantum modeling languages for specifying QSAs, Carlos et al. [S32] have developed Q-UML - an extension to classical UML (Unified Modeling Language) – to support structural and behavioral representation of quantum search algorithms \cite{UML}. Specifically, considering the (co-) design and implementation challenges of QSAs, the role of architectural modeling becomes pivotal to provide a software blue-print model that acts as a bridge between architectural requirements and their implementations, as in Figure \ref{QSA}. Architectural models essentially becomes the driving artifact in the context of  model-driven architecting, where architectural models and model transformation can be exploited for model-based implementation and validation of the system \cite{RQ1-11}. 
To systematically classify, analyse and compare architectural modeling or description languages, some frameworks have been developed that provide a criteria-driven analysis of architectural modeling \cite{RQ1-5}\cite{TSEADL}. These evaluation criteria can be generally classified into three main types, each type exploring the role of modeling notations to support (i) \textbf{architectural specifications} (e.g., architectural representation, architectural structure, syntax, and semantics and, analysing static and dynamic nature of the architectures), (ii) \textbf{quality attributes} (e.g., extension, customization, interoperability of the notations), (iii) \textbf{architectural process} (architectural requirements, implementation, validation). The focus of this RQ is architectural representation, not quality attributes of modeling notations, therefore, we mainly focus on aspects of architectural representation and support for architectural process (Figure \ref{QSA}) with the help of Table \ref{tab:ModelingNotation}. Table \ref{tab:ModelingNotation} acts as a structured catalogue to summarise the following information to answer this question.

\begin{table}[!t]\footnotesize
\centering
\scriptsize
\caption{Summary View of of Modeling Notations, Modeling Artifacts, and Lifecycle Support. (AR = Architectural Requirements, AD = Architectural Design, AI = Architectural Implementation, AE = Architectural Evaluation, AT= Architectural Deployment)}
\label{tab:ModelingNotation}
\begin{tabular}{|cll|lllll|}
\hline
\multicolumn{1}{|c|}{\textbf{Study ID}} & \multicolumn{1}{l|}{\textbf{Modeling Notation}} & \textbf{Modeling Artifact} & \multicolumn{5}{l|}{\textbf{Process Support}} \\ \hline
\multicolumn{3}{|l|}{\textbf{}} & \multicolumn{1}{l|}{\textbf{AR}} & \multicolumn{1}{l|}{\textbf{AD}} & \multicolumn{1}{l|}{\textbf{AI}} & \multicolumn{1}{l|}{\textbf{AE}} & \textbf{AD} \\ \hline
\multicolumn{1}{|l|}{S1} & \multicolumn{1}{l|}{Box and arrows} & Component diagram & \multicolumn{1}{l|}{} & \multicolumn{1}{l|}{\checkmark} & \multicolumn{1}{l|}{} & \multicolumn{1}{l|}{} &  \\ \hline
\multicolumn{1}{|l|}{S2} & \multicolumn{1}{l|}{Graph-based model} & State graph & \multicolumn{1}{l|}{} & \multicolumn{1}{l|}{\checkmark} & \multicolumn{1}{l|}{\checkmark} & \multicolumn{1}{l|}{} &  \\ \hline
\multicolumn{1}{|l|}{S4} & \multicolumn{1}{l|}{UML} &Class diagram & \multicolumn{1}{l|}{} & \multicolumn{1}{l|}{\checkmark} & \multicolumn{1}{l|}{\checkmark} & \multicolumn{1}{l|}{} & \checkmark \\ \hline
\multicolumn{1}{|l|}{S5} & \multicolumn{1}{l|}{Graph-based model} & Process flow model & \multicolumn{1}{l|}{\checkmark} & \multicolumn{1}{l|}{\checkmark} & \multicolumn{1}{l|}{} & \multicolumn{1}{l|}{} &  \\ \hline
\multicolumn{1}{|l|}{S6} & \multicolumn{1}{l|}{Box and arrows} & \begin{tabular}[c]{@{}l@{}}State transition diagram\\    \\ \end{tabular} & \multicolumn{1}{l|}{} & \multicolumn{1}{l|}{\checkmark} & \multicolumn{1}{l|}{} & \multicolumn{1}{l|}{} & \checkmark \\ \hline
\multicolumn{1}{|l|}{S7} & \multicolumn{1}{l|}{Graph-based model} & State graph & \multicolumn{1}{l|}{} & \multicolumn{1}{l|}{\checkmark} & \multicolumn{1}{l|}{} & \multicolumn{1}{l|}{} &  \\ \hline
\multicolumn{1}{|l|}{S9} & \multicolumn{1}{l|}{UML} & State transition diagram & \multicolumn{1}{l|}{\checkmark} & \multicolumn{1}{l|}{\checkmark} & \multicolumn{1}{l|}{} & \multicolumn{1}{l|}{} &  \\ \hline
\multicolumn{1}{|l|}{S14} & \multicolumn{1}{l|}{Box and arrows} & Component diagram & \multicolumn{1}{l|}{} & \multicolumn{1}{l|}{} & \multicolumn{1}{l|}{\checkmark} & \multicolumn{1}{l|}{\checkmark} &  \\ \hline
\multicolumn{1}{|l|}{S21} & \multicolumn{1}{l|}{\begin{tabular}[c]{@{}l@{}}Box and arrows\\    \\ Graph-based model\end{tabular}} & \begin{tabular}[c]{@{}l@{}}\\    \\ State graph\end{tabular} & \multicolumn{1}{l|}{\checkmark} & \multicolumn{1}{l|}{\checkmark} & \multicolumn{1}{l|}{} & \multicolumn{1}{l|}{} &  \\ \hline
\multicolumn{1}{|l|}{S22} & \multicolumn{1}{l|}{Box and arrows} & Component diagram & \multicolumn{1}{l|}{} & \multicolumn{1}{l|}{\checkmark} & \multicolumn{1}{l|}{} & \multicolumn{1}{l|}{\checkmark} &  \\ \hline
\multicolumn{1}{|l|}{S25} & \multicolumn{1}{l|}{Graph-based model} & State graph & \multicolumn{1}{l|}{} & \multicolumn{1}{l|}{\checkmark} & \multicolumn{1}{l|}{} & \multicolumn{1}{l|}{} & \checkmark \\ \hline
\multicolumn{1}{|l|}{S27} & \multicolumn{1}{l|}{Graph-based model} & Process flow model & \multicolumn{1}{l|}{} & \multicolumn{1}{l|}{\checkmark} & \multicolumn{1}{l|}{\checkmark} & \multicolumn{1}{l|}{\checkmark} &  \\ \hline
\multicolumn{1}{|l|}{S28} & \multicolumn{1}{l|}{\begin{tabular}[c]{@{}l@{}}Box and arrows\\    \\ Graph-based model\end{tabular}} & \begin{tabular}[c]{@{}l@{}}\\    \\ State graph\end{tabular} & \multicolumn{1}{l|}{} & \multicolumn{1}{l|}{\checkmark} & \multicolumn{1}{l|}{\checkmark} & \multicolumn{1}{l|}{} &  \\ \hline
\multicolumn{1}{|l|}{S31} & \multicolumn{1}{l|}{Graph-based model} & Process flow model & \multicolumn{1}{l|}{} & \multicolumn{1}{l|}{\checkmark} & \multicolumn{1}{l|}{} & \multicolumn{1}{l|}{} &  \\ \hline
\multicolumn{1}{|l|}{S32} & \multicolumn{1}{l|}{UML (Q-UML)} & \begin{tabular}[c]{@{}l@{}}Class diagram \\    \\ Sequence diagram\end{tabular} & \multicolumn{1}{l|}{\checkmark} & \multicolumn{1}{l|}{\checkmark} & \multicolumn{1}{l|}{} & \multicolumn{1}{l|}{} &  \\ \hline
\multicolumn{1}{|l|}{S33} & \multicolumn{1}{l|}{Box and arrows} & \begin{tabular}[c]{@{}l@{}}\\    \\ Component diagram\end{tabular} & \multicolumn{1}{l|}{} & \multicolumn{1}{l|}{\checkmark} & \multicolumn{1}{l|}{\checkmark} & \multicolumn{1}{l|}{} & \checkmark \\ \hline
\end{tabular}
\end{table}

\textit{\textbf{Available evidence}} reflects the published research, that provides details of the modeling notation for QSAs [S32].

\textit{\textbf{Modeling notation}} represents a specific method or technique that is being used to represent the model for QSA. For example, the Q-UML solution provided by Carlos et al. [S32] is an extension of the UML for structural and behavioral representation of the QSA. In addition to the extensions of already existing modeling notations (i.e., QSA specific tailoring), conventional notations such as graph-based models or box and arrow structures have been exploited to specify the structure and semantics of QSAs [S16][S27]. For example, Killoran et al. [S27] exploits graph-based models to represent modules of code to implement the quantum software. Specifically, in graph-based modeling the modules of source code are represented as graph nodes (computational elements and data stores), whereas graph edges represent the interconnection the code modules. This means that \textbf{TransactionCommit} module (\textbf{node\_1}) transfers control to \textbf{Update TransactionRecord} module (\textbf{node\_2}) via \textbf{commit} connector (\textbf{edge\_A}) in architectural graph for quantum software.

\textit{\textbf{Modeling artifact}} represents a specific artifact (i.e., visual diagram, model) to represent an instance of the architectural model. For example, Carlos et al. [S32] used UML class diagram is being used to represent the structure, whereas UML sequence diagrams are used to represent the behavior of the quantum search algorithm.

\textit{\textbf{Architectural process support}} needs modeling notation (and its underlying artifacts) to support specific activities in the architectural process from RQ2.1. For example, Q-UML presents class and sequence diagrams to (i) model requirements and (ii) specify structural representation and execution flow of the quantum search design. The proposed solution Q-UML does not provide support for other architecting activities such as architectural implementation or evaluation.

Table \ref{tab:ModelingNotation} summarises the core findings of RQ2.2 to streamline most adopted modeling notations, the artifacts being used to model the QSAs, and their impacts on architectural process. We can conclude that most prominent modeling notations can be broadly classified into three main types as UML profiles and extensions such as [S4, S9, S32] (3 studies), graph-based models including [S2, S5, S7, S21, S25, S27, S28, S31] (8 studies), and box and arrow notations including [S1, S6, S14,  S16, S19, S21, S22, S28, S33] (9 studies). Some of the most used state transition diagrams, state graph, and process flow models diagram. In the context of architectural process support, existing modeling notations are primarily focused on supporting architectural requirements [S5, S9, S16, S21, S32] (05 studies), design [S1, S2, S4, S5, S6, S7, S9, S16, S19, S21, S22, S25, S27, S28, S31, S32, S33] (17 studies) and implementation phases [S2, S4, S14, S19, S27, S28, S33] (7 studies), whereas there is much less support for life-cycle activities like architectural evaluation [S14, S22, S27] (3 studies) and deployment [S4, S6, S25, S33] (4 studies). Modeling notations are fundamental to the creation of architectural design models that provide foundations for architectural implementation \cite{UML}. In the context of this research, models can facilitate other architectural aspects including but not limited to design decisions (patterns and styles that promote reuse) and tools that support customization, human decision support, and automation, detailed in subsequent sections of this paper.

\begin{tcolorbox}[colback=gray!5!white,colframe=gray!75!black,title=Key Findings of RQ2.2]
\textbf{Finding 9}: Modeling notations to specify quantum software architectures primarily rely on \textit{ box and arrow notations} (having component diagrams) and \textit{graph-based models } (having state graph) to represent the structures and behavior of quantum software under design. Unlike conventional software architectures that mostly exploit UML notations (often considered as a defacto approach for software design), there is much less evidence on UML-based modeling quantum software architectures

\textbf{Finding 10}: It appears that there is a need for \textit{architectural description languages} and \textit{UML profiles} that can be helpful to leverage existing tools, frameworks, and architectural knowledge to empower the role of designers and architects to model, develop, and evolve quantum software based on re-usability and (semi-) automation.
\end{tcolorbox}

\subsection{Architecture design patterns (RQ2.3)} \label{RQ2.3}
To answer RQ2.3, we identified a total of six quantum software architecture patterns discussed in (n=17, 50\%) studies. In design or architectural context, patterns represent reusable design knowledge, referred to as best practices and concentrated wisdom of designers to address recurring challenges of software development. For example, to address the challenges of system structuring and deployment the layered architecture pattern helps architects to organise software-intensive systems and applications into various layers, each dedicated to different concerns such as data management, user interfacing and computations [S1, S18]. A collection of patterns formally or informally organized into a sequence, results in architectural pattern languages \cite{RQ1-7}. The focus of this study is individual patterns rather than pattern languages. The set of identified quantum software architecture patterns is presented in Table \ref{tab:QSEpatterns}. The most recurring design patterns discussed in the 18 primary studies are \textit{layered (n=8, 24\%)} and \textit{pipe and filter architecture (n=5, 15\%}) patterns. The other patterns having low frequency of occurrence are \textit{(composite design, prototype design, recursive containment and two-qubit gate)}. In the following text, we briefly describe the example of a \textit{layered pattern} for the general-purpose microarchitecture of quantum software [S3]. Generally, the \textit{layered pattern} architecture of quantum software mainly consists of several properties that we also need to estimate. These properties include appropriate instruction length, pipeline depth (for parallel quantum gates), and multiple control channels per single instruction. These properties help to construct the basic blocks of quantum software, such as the timing control unit and the microcode instruction set of the overall system. According to our results, the second most frequently reported pattern used for designing quantum software is \textit{pipe and filter}. Killoran et al. [S27] proposed an open-source quantum programming architecture (i.e., Strawberry Fields) based on pipe and filter patterns. The elements of the proposed architecture are organized as the front-end and the back-end. The front-end layer consists of interactive server, application, field API, and quantum programming language components, and the back-end components include a quantum processor and simulator. Both layers communicate through the compiler engine.  Our results indicate that the patterns for quantum software are similar to other types of software (e.g., monolithic based architecture, services-oriented based architecture, microservices-based architecture). However, these patterns deal with a series of instructions that need to be executed on quantum processors.

\begin{tcolorbox}[colback=gray!5!white,colframe=gray!75!black,title=Key Findings of RQ2.3]
\textbf{Finding 11}: \textit{Layered} and \textit{pipe and filter} patterns are identified as the most recurring quantum software architecture patterns. However, these are generic or classical patterns that can be used to design any software system. To this end, further research efforts are required to explore and propose new patterns to particularly focus on quantum computing attributes (e.g. superposition and quantum entanglement) and facilitate the architecture of quantum software systems.
\end{tcolorbox}

\begin{table}[!t]\footnotesize
\centering
\scriptsize
\caption{Quantum software architecture design patterns}
\label{tab:QSEpatterns}
\begin{tabular}{|p{0.65\textwidth}|l|}
\hline
\textbf{Pattern Name} & \textbf{Study IDs} \\ \hline
Layered pattern & S3, S5, S9, S14, S18, S26, S28, S29 \\ \hline
Pipe and filter pattern & S2, S20, S21, S27, S31 \\ \hline
Composite design pattern & S4 \\ \hline
Prototype design pattern & S24 \\ \hline
Recursive containment & S9 \\ \hline
Two-qubit gate pattern & S20 \\ \hline
\end{tabular}
\end{table}
\subsection{Architecture tools and frameworks (RQ2.4)} \label{RQ2.4}
RQ2.4 is developed to identify tools and frameworks used to support the architecting activities discussed in Section \ref{RQ2.1}. We explored the selected primary studies and noticed that only \textit{(n=11, 32\%)} studies discussed architectural tools and frameworks (see Table \ref{tab:Tools}). The tools and frameworks provide semi- or fully automated solutions to perform architecting activities. Tools broadly refer to software solutions that automate, enhance, or customise process activities. On the other hand, a framework is a set of tools used to perform a bunch of activities, e.g., designing, implementation, and documentation. Each identified tool and framework is interpreted based on the following five criteria \cite{sajjad2018classification}, as listed in Table \ref{tab:Tools}). \textbf{Source type} refers to the type as open source (OS) or close source (CS). In open source, the copyright holders grant the user permissions to study, use or update the tool, framework or system. \textbf{Input instructions} are the instructions provided to execute the logic. The instruction types are categorised as high-level (HL), quantum instruction (QI), and mathematical variables (MV). \textbf{Output} are the type of post execution findings and categorised as quantum source code (QSC), quantum algorithm (QA), and simulation findings (SF). \textbf{Automation level} refers to the automation level of the tool or framework. Automation could be fully-automated (FA), semi-automated (SA), or non-automated (NA). \textbf{Evaluation} refers to the performance assessment of a particular tool and framework. Evaluation could be explicit (EX) or implicit (IM). Implicit means that tool or framework is partially evaluated or few of the components are empirically assessed.


\begin{table*}[t]\footnotesize
 \centering
 \scriptsize
  \caption{List of identified tools}
\label{tab:Tools}
 \begin{tabular}{|p{3.5cm}|p{1cm}|p{1.5cm}|p{1cm}|p{1.6cm}|p{1.5cm}|p{1cm}|}
 \hline
 \textbf{Tool/Framework}&\textbf{Source Type}&\textbf{Input Instructions}&\textbf{Output}& \textbf{Automation level}&\textbf{Evaluation}&\textbf{Study}\\\hline
 XACC (eXtreme-scale   Accelerator) & CS & HL & QSC & FA & EX & S4 \\ \hline
Link layer & CS & QI & SF & FA & EX & S6 \\ \hline
Auto   E/E framework & OS & MV & SF & SA & IM & S9 \\ \hline
eQASM & CS & QI & QA & SA & EX & S14 \\ \hline
JKQ   (tool set) & OS & HL & SF & FA & EX & S16 \\ \hline
Kwant & OS & MV & SF & FA & EX & S17 \\ \hline
JKQ DDSIM & OS & HL & SF & FA & EX & S19 \\ \hline
QuNetSim & OS & HL & SF & SA & IM & S25 \\ \hline
Strawberry   fields & OS & HL & SF & FA & EX & S27 \\ \hline
qcor & OS & HL & SF & FA & EX & S28 \\ \hline
GH-QPL & CS & HL & QSC & SA & IM & S33 \\ \hline
 \end{tabular}

\end{table*}

\begin{figure}[!t]
 \centering
 \includegraphics[width=\linewidth]{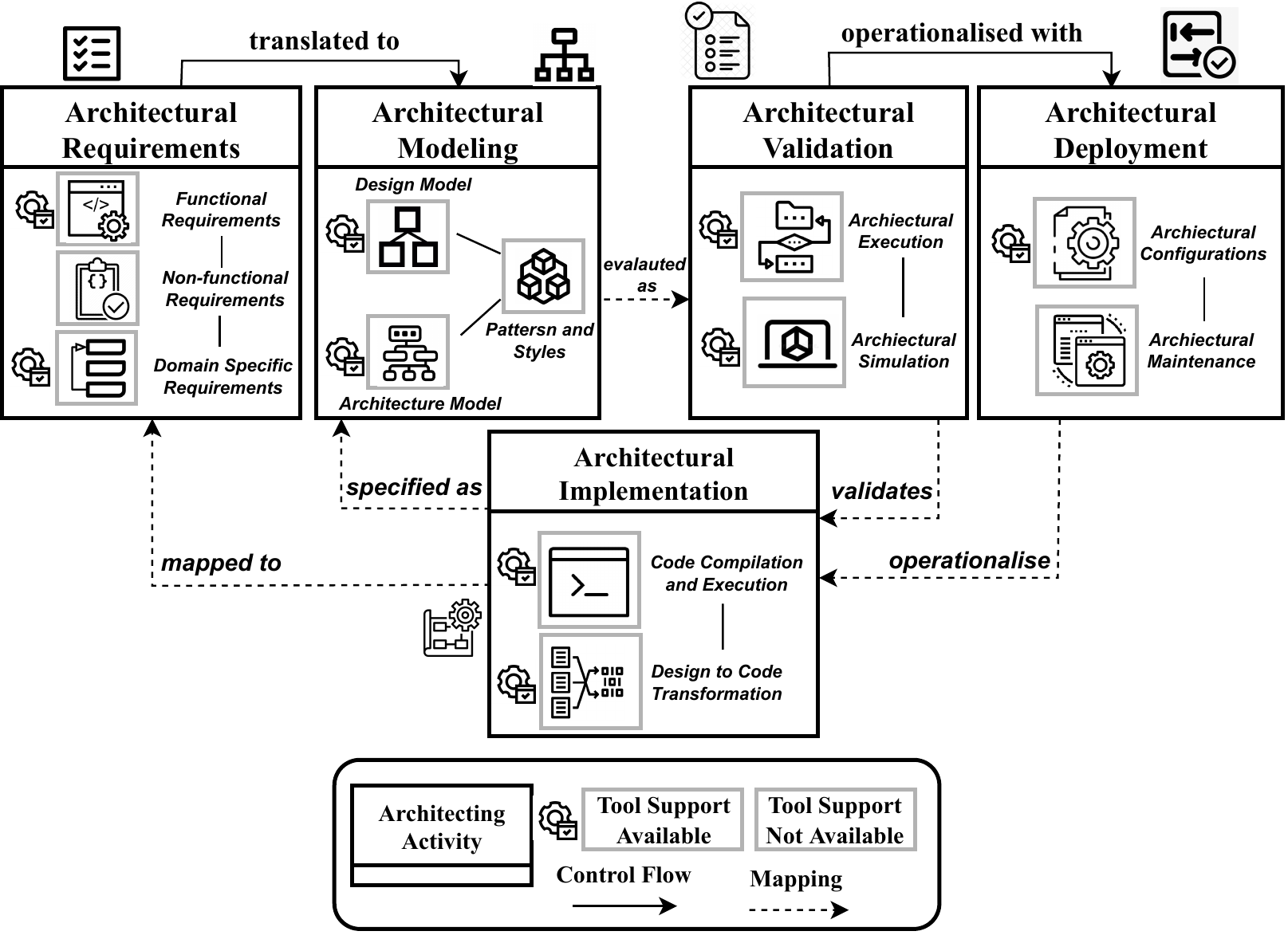}
 	\caption{Tool support for architecting activities}
	\label{Fig:Toolsupport}
\end{figure}

The results given in Table \ref{tab:Tools} reveal that \textit{(n=7, 64\%)} tools and frameworks are open source (OS). Similarly, \textit{(n=7, 64\%)} tools and frameworks accept  input code in high-level (HL) programming format (i.e instructions that are more or less independent of a specific type of computer). Moreover, \textit{(n=8, 73\%)} tools and frameworks simulate the high-level input instructions and give the output based on the simulation findings (SF). We further noticed that \textit{(n=7, 64\%)} tools and frameworks are fully-automated (FA) and \textit{(n=8, 73\%)} are explicitly (EX) evaluated based on their performance. The visualisation and summary of the results on tool support are provided in Figure \ref{Fig:Toolsupport} and Table \ref{tab:ToolsFramework}.

Finally, the identified tools and frameworks are classified with respect to their contribution across the architectural process activities reported in Section \ref{RQ2.1}. Thematic analysis approach discussed in Section \ref{Synthesis} is followed to categorise the identified tools and frameworks and present the toolchain. It should be noted that a specific tool or framework might contribute to more than one architecting activities and we consider them across multiple activities (see Table \ref{tab:ToolsFramework}). 





The core architecting activities with respect to the tools and frameworks support are subsequently discussed:

\textit{Architectural requirements:} We explored the selected primary studies and identified a single framework that focuses on \textit{architectural requirements} (see Table \ref{tab:ToolsFramework})  [S9]. Lan et al. [S9], proposed a quantum computing based architectural framework to minimize the gap between the functional domains and meet the requirements of the open electrical and electronic automotive embedded systems. Architectural requirements is a less focus activity with respect to tools and frameworks and the reason might be that quantum software architecture field is in the evolution phase and still the architectural requirements activities do not have tool based automation and customization support.  

\textit{Architectural implementation:} We identified that a total of six tools and frameworks contributed to the \textit{architectural implementation} activity (see Table \ref{tab:ToolsFramework}). More narrow, these tools and frameworks explicitly focus on the \textit{code compilation} and \textit{design to code transformation} sub-activities (see Figure \ref{Fig:Toolsupport}). The power of quantum computer could only be realised by implementing quantum algorithms to control the hardware devices, improve the performance and verify the quantum attributes \cite{WinNTRoberto}. Therefore, researchers and practitioners are rushing to develop strategies, tools, frameworks and guidelines to implement algorithms in a simple and efficient way. For example, XACC (eXtreme-scale ACCelerator) provides interfaces to enhance hybrid compilation of programs developed both in quantum and classical programming languages [S4]. XACC programming framework is designed in a manner that it is entirely independent of selected language, computational model and hardware. The implementation tools and frameworks instantly assist in realizing the real-world computation benefits of quantum computers and increase its application across various industrial domains.

\textit{Architectural modeling:} We noticed that only two  \textit{architectural modeling } tools and frameworks are developed, which explicitly address \textit{design model} and \textit{architecture model} sub-activities [S27, S28] (see Figure \ref{Fig:Toolsupport}). Modeling activities performed to develop the overall architecture, which acts as a blueprint for the implementation. The quantum software engineering field is still undeveloped, and it is important to create high-level modeling abstractions for classical software engineers to understand and model the quantum programs. For example, Strawberry Fields is an open source architectural framework developed to design and optimize the software systems for photonic quantum computers [S27]. Strawberry Fields has built-in engine to convert the code developed in domain specific programming language (blackbird) and run using the photonic quantum computers.

\begin{table}[!t]\footnotesize
\centering
\scriptsize
\caption{Summary view of tools and frameworks across architecting activities (AR = Architectural Requirements, AM = Architectural Modeling, AI = Architectural Implementation, AV = Architectural Validation, AD= Architectural Deployment)}
\label{tab:ToolsFramework}
\begin{tabular}{|cll|lllll|}
\hline
\multicolumn{1}{|c|}{\textbf{Study ID}} & \multicolumn{1}{l|}{\textbf{Tool Name}} & \textbf{Tool Focus} & \multicolumn{5}{l|}{\textbf{Process Support}} \\ \hline
\multicolumn{3}{|l|}{\textbf{}} & \multicolumn{1}{l|}{\textbf{AR}} & \multicolumn{1}{l|}{\textbf{AM}} & \multicolumn{1}{l|}{\textbf{AI}} & \multicolumn{1}{l|}{\textbf{AV}} & \textbf{AD} \\ \hline

\multicolumn{1}{|l|}{S4} & \multicolumn{1}{l|}{xACC} & Code compilation & \multicolumn{1}{l|}{} & \multicolumn{1}{l|}{} & \multicolumn{1}{l|}{\checkmark} & \multicolumn{1}{l|}{} &  \\ \hline

\multicolumn{1}{|l|}{S9} & \multicolumn{1}{l|}{Auto.E/E Framework} & Requirements & \multicolumn{1}{l|}{\checkmark} & \multicolumn{1}{l|}{} & \multicolumn{1}{l|}{} & \multicolumn{1}{l|}{} &  \\ \hline

\multicolumn{1}{|l|}{S27} & \multicolumn{1}{l|}{Strawberry fields} & Domain modeling & \multicolumn{1}{l|}{} & \multicolumn{1}{l|}{\checkmark} & \multicolumn{1}{l|}{} & \multicolumn{1}{l|}{} &  \\ \hline

\multicolumn{1}{|l|}{S28} & \multicolumn{1}{l|}{qCOR} &Design & \multicolumn{1}{l|}{} & \multicolumn{1}{l|}{\checkmark} & \multicolumn{1}{l|}{\checkmark} & \multicolumn{1}{l|}{} & \\ \hline

\multicolumn{1}{|l|}{S14} & \multicolumn{1}{l|}{eQASIM} & Program flow and execution & \multicolumn{1}{l|}{} & \multicolumn{1}{l|}{} & \multicolumn{1}{l|}{\checkmark} & \multicolumn{1}{l|}{} &  \\ \hline

\multicolumn{1}{|l|}{S16} & \multicolumn{1}{l|}{JKQ} & Code compilation & \multicolumn{1}{l|}{} & \multicolumn{1}{l|}{} & \multicolumn{1}{l|}{\checkmark} & \multicolumn{1}{l|}{\checkmark} &  \\ \hline

\multicolumn{1}{|l|}{S19} & \multicolumn{1}{l|}{JKQ DDSIM} & Simulation, compilation & \multicolumn{1}{l|}{} & \multicolumn{1}{l|}{} & \multicolumn{1}{l|}{\checkmark} & \multicolumn{1}{l|}{} &  \\ \hline

\multicolumn{1}{|l|}{S33} & \multicolumn{1}{l|}{GH-QPL} &Translation and compilation & \multicolumn{1}{l|}{} & \multicolumn{1}{l|}{} & \multicolumn{1}{l|}{\checkmark} & \multicolumn{1}{l|}{} &  \\ \hline

\multicolumn{1}{|l|}{S6} & \multicolumn{1}{l|}{LinkLayer} & Quantum communication & \multicolumn{1}{l|}{} & \multicolumn{1}{l|}{} & \multicolumn{1}{l|}{} & \multicolumn{1}{l|}{} & \checkmark \\ \hline

\multicolumn{1}{|l|}{S17} & \multicolumn{1}{l|}{Kwant} &Simulation & \multicolumn{1}{l|}{} & \multicolumn{1}{l|}{} & \multicolumn{1}{l|}{} & \multicolumn{1}{l|}{\checkmark} &  \\ \hline

\multicolumn{1}{|l|}{S19} & \multicolumn{1}{l|}{JKQ DDSIM} &Simulation & \multicolumn{1}{l|}{} & \multicolumn{1}{l|}{} & \multicolumn{1}{l|}{} & \multicolumn{1}{l|}{\checkmark} &  \\ \hline

\multicolumn{1}{|l|}{S25} & \multicolumn{1}{l|}{QuNetSim} &Simulation& \multicolumn{1}{l|}{} & \multicolumn{1}{l|}{} & \multicolumn{1}{l|}{} & \multicolumn{1}{l|}{\checkmark} &  \\ \hline

\end{tabular}
\end{table}

\textit{Architectural deployment:} Finally, we noticed that only one framework focuses on deployment activities i.e., link layer [S6]. It is developed for quantum communication that improves the entanglement attributes between quantum computers into robust and well defined services. Additionally, strategies for network scheduling are developed to evaluate the protocol performance with respect to different use cases. Architectural deployment is a slightly less focused activity and in near term the tools to automate the deployment activities will be demanding need. 

\begin{tcolorbox}[colback=gray!5!white,colframe=gray!75!black,title=Key Findings of RQ2.4]
\textbf{Finding 12}: The identified tools and frameworks are categorise based on the five core attributes namely \textit{(i) source type, (ii) inputs, (iii) outputs, (iv) automation,} and \textit{(v) evaluation level}.

\textbf{Finding 13}: The identified tools and frameworks are mapped across the architecting activities and presented as a toolchain (see Figure \ref{Fig:Toolsupport}). \textit{Architectural implementation} is identified as the most common activity with respect to tools and frameworks. We noticed that six tools and frameworks (n=6, 55\%) are developed to automate and customise the architectural implementation activities.  
\end{tcolorbox}

\subsection{Architecture challenges} \label{RQ2.5}
The selected primary studies are explored to identify the key challenges of quantum software architecture (RQ2.5). We found that only \textit{(n=16, 47\%)} primary studies reported the architecture challenging factors. The identified challenges are further classified across four core themes: \textit{quantum data transmission and security}, \textit{process-centric architecting}, \textit{architectural tools and technological support}, and \textit{architecting knowledge and expertise}. The thematic analysis approach discussed in Section \ref{Synthesis} is followed to systematically identify the most common themes of the challenging factors (see Figure \ref{Fig:Challengesthemes}). For fine-grained analysis, the main themes (core categories) and sub-themes (challenging factors) are presented in Figure \ref{Fig:Challengesthemes} and explicitly discussed below:

\subsubsection{Quantum data transmission and security}
\begin{figure}
 \centering
 \includegraphics[width=\linewidth]{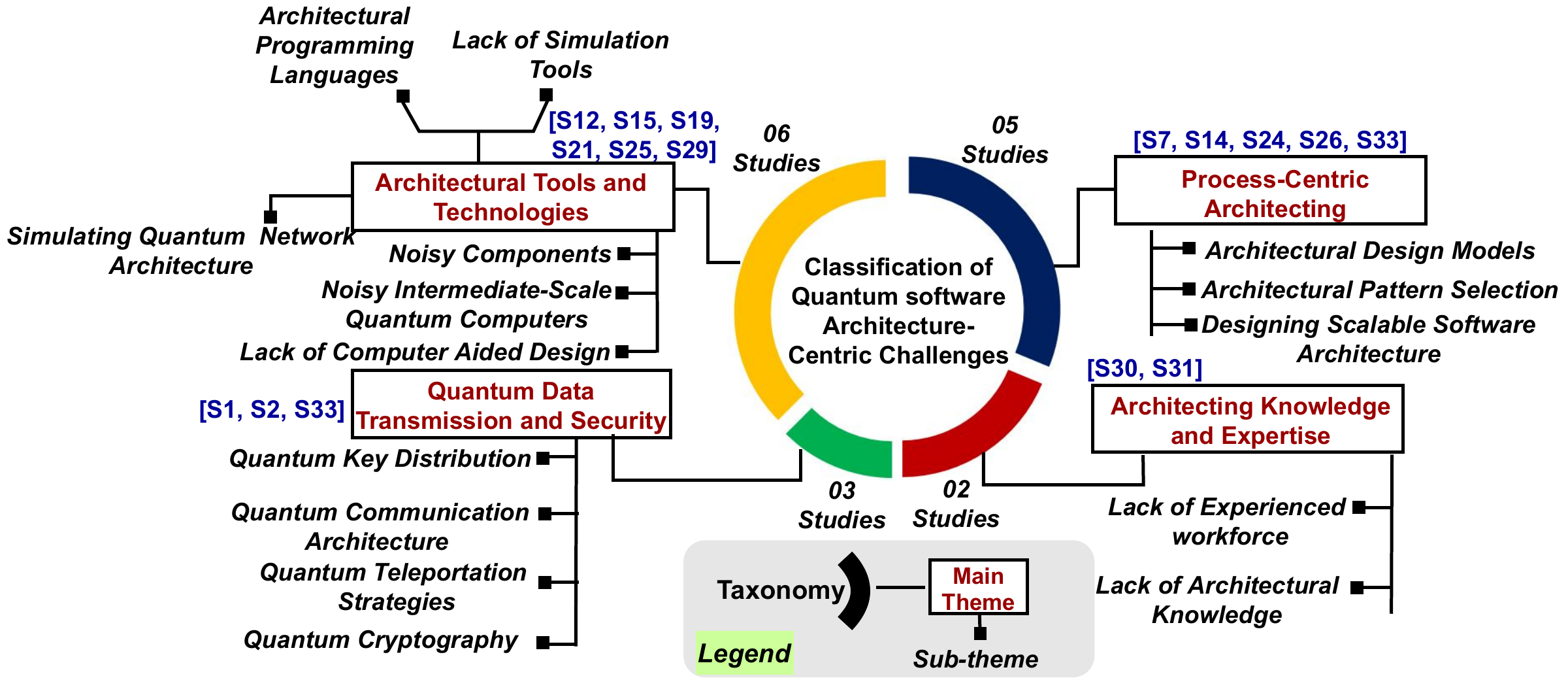}
 	\caption{Thematic classification of identified challenges}
	\label{Fig:Challengesthemes}
\end{figure}

This theme covers the challenging factors related to the security of network architecture developed for quantum data transmission. We identified a total of four sub-themes (challenging factors) related to the security of quantum network architecture (see Figure \ref{Fig:Challengesthemes}). The identified challenging factors are thoroughly discussed as follow:

\textit{Quantum key distribution (QKD):} The quantum key distribution (QKD) approach is used to develop the ultra-secure network for quantum data transmission [S1]. QKD involves sending the encrypted data and decryption keys over quantum network in qubit state.  However, the existing QKD systems are designed to work on the single link quantum network and becomes challenging to operate across multiple networks where the system design and protocols get more complex [S1,S2]. It is evident that there is a strong need of QKD architecture that could deploy across multiple networks for transmitting secure quantum data.\\

\textit{Quantum communication architecture: } Architecting  a  quantum  network  is  challenging  with  respect  to  communication  perspectives. Quantum network architecture is distinct to classical because of quantum attributes including superposition, entanglement, and quantum measurement [S33]. These attributes brings significant constraints to design the quantum communication architecture. In classical communication, the data bits used to convey the message. In contrast, the qubits are used to transmit the data over quantum communication channel, however; developing a quantum communication architecture needs a major paradigm shift to consider the characteristics of quantum mechanics [S33]. The open-source community should join the efforts to design and fabricate the quantum communication architecture models and interfaces.

\textit{Quantum teleportation strategies: } Techniques used to transfer quantum information between sender and receiver is called quantum teleportation. Teleportation in science fiction refers to transfer a physical object from location A to B; however, in quantum computing it is used to transfer the Qubits. It has pivotal role in the continuing progress of quantum communication, and quantum networks. However, teleportation is a major challenge in present day quantum computing science because of lack of teleportation protocols, strategies and techniques. Qubits transmission across multiple nodes and computation in the cloud domain is only possible by using the quantum teleportation strategies [S33]. There is a strong need for teleportation protocols and strategies that could reshape the quantum teleportation process.

\textit{Quantum cryptography: } Practically, quantum cryptography is in its infancy because of data transmission rates and processing limitations. These issues are complicated and challenging to tackle as the high-quality single photons for long-distance required low transmission loss rates. It increases the technological cost of quantum cryptography as compared to the classical. Similarly, developing a sharing infrastructure for secure data encryption and decryption is a significant challenge for quantum cryptography [S33]. The effective encryption and decryption solution is possible by introducing the intermediate node between the sender and receiver. Presently, tackling quantum cryptography challenges is complex, and world-leading technology giants are racing to propose effective solutions. 

\subsubsection{Process-centric architecting}
This theme is developed to categorize the key challenging factors (sub-themes) that could impact the design process of quantum software architecture. Following is the detail description of each selected challenge that covers the process-centric theme. 

\textit{Architectural design models: } There is a lack of models for designing quantum software architectures. The existing models are simplified extended versions of classical modeling approaches and do not explicitly cover the quantum properties including superposition, interference, and entanglement [S7, S24]. The unavailability of particular quantum software design models make it hard to design the system architecture. The expectations to consider quantum computing as alternative to classical increased exponentially [S7, S24]. Consequently, it becomes important to propose rigorous design models in advance for architecting quantum software systems.

\textit{Architectural pattern selection: } Architectural pattern is a common and reusable solution for generally occurring architectural problems. Selecting an appropriate architecture pattern for a specific quantum problem is a challenging feat. The multi-criteria decision making (MCDM) model could be a best solution to choose a right pattern for right problem [S26]. MCDM model provides a platform to tackle the commonly occurred quantum architectural problems.

\textit{Designing scalable quantum software architecture: } Scalable systems refer to the information processing concept where a complex system could be developed using the basic building blocks. In quantum architectural scalability, the qubits properties improve or remains consistent when they are extended across multi-qubits systems [S14, S33]. However, architectural scalability also needs to consider the Qubits operations with specific timing, in time instructions fetching and processing to ensure that desired operations are accurately performed [S14]. It is hard to live up the real-world promises and supremacy of quantum computers without architectural scalability [S33].

\subsubsection{Architectural Tools and Technologies}
The tools and technologies theme is developed to classify the challenges related to the technical support for architecting activities. In-depth discussion of these challenges is provided as follows:  

\textit{Noisy Components:} Constructing a scalable quantum computer is challenging due to environmental interaction noise that could destroy its highly fragile components [S12]. Environmental interaction noise generated because of control devices and heat, which can seriously disturb the qubits superposition state and cause computational errors. The robust statistical and mathematical models to estimate the noise impact can significantly improve the computation process and protect the superposition state [S12].

\textit{Noisy Intermediate-Scale Quantum (NISQ) computers:} It will take decades of research to realize the fault-tolerant quantum computer for solving the wide range real-world problems [S15]. However, the concept of noisy intermediate-scale quantum (NISQ) computer already exists, which contains fifty to a few hundred Qubits but is not smart enough to continuously perform fault-free computations [S15, S21]. The term noisy is used because the present day quantum processors are not sophisticated enough to cope with the environmental impacts, which cause to lose the quantum coherence. Experimental interest is expected and demanded in designing quantum software and hardware architectures to process and execute a large number of error-free Qubits. Transition to quantum computing or more specifically adopting the quantum hardware and quantum computing platforms requires financial investments as well as human skills to manage quantum resources. The PISQ (Perfect Intermediate Scale Quantum) enables the development of new software applications by developing algorithms and evaluating them on quantum simulators that can be executed on existing computing platforms \cite{PISQ}. Solutions like PISQ may not be long term solutions to support quantum software, but such solutions allow research and development of quantum logic via simulations that can be deployed and executed on non-quantum computing platforms.

\textit{Lack of computer-aided design (CAD) tools:} Computer-aided design tools enable the development, change, and optimization of the architecture design process. These tools are significantly important for developing nanoscale quantum software architectures [S19].  Research to automate and optimise the design approaches for quantum software systems is boosting; however, there is a considerable coordination gap between the CAD and quantum computing community [S19]. Consequently, various proposed CAD tools are failed to achieve the core architectural objectives.

\textit{Simulating quantum networks architecture:} The quantum internet is defined to transmit quantum data, which is a network architecture of multiple devices and software tools. The concept of a quantum internet is still not in practice, and development efforts are being made to shape it practically. To analyze network protocols, it is important to assess their significance using different simulation tools [S25]. However, limited studies discussed such tools for evaluating quantum network protocols and there is a strong need for advanced simulation tools.

\textit{Architectural programming languages:} Quantum architectural programming language should provide all the required abstractions both to quantum physicists and algorithm designers. The existing  languages are not rich enough to consider for future high-number Qubits algorithms [S29]. They are still unpredictable for complex quantum problems. In the future, the architectural languages should support high-level abstractions for developing and deploying advance  algorithms based on quantum superposition and entanglement. Quantum programming languages and frameworks provided by technology giants (e.g., Qiskit by IBM, and Q\# by Google) enable software developers to implement QSA as quantum source code that can be executed or implemented on quantum computing platforms. However, the results of an exploratory study show that (i) mined quantum source code repositories available on GitHub and (ii) interviewed quantum code developers suggests that beyond the industry led projects, adoption and applicability of quantum programming in developers’ community is still limited \cite{QuantProg}. The study also highlights that the current generation of software developers, while implementing quantum code, face a multitude of challenges that range from quantum program comprehension to source code analysis, manipulation, and testing \cite{wang2022qusbt}.

\textit{Lack of simulation tools:} The lack of simulation tools is considered a major barrier for quantum software architecture research. The need of simulation tools escalates for large-scale practical and reliable measurements [S29]. Generally, the architects are interested in knowing how fast the architecture works for a specific application, which types of operations it can perform, and what would be the reliability level of its results? These questions could possibly be answered by proposing particular simulators for quantum software architecture [S29]. OpenQL provides a quantum programming language and its associated quantum compiler to develop and execute quantum source code. OpenQL also produces quantum assembly code that is technology independent and can be simulated using QX Quantum Computer Simulator \cite{OpenQL}.

\subsubsection{Architecting knowledge and expertise}
Designing a real-world quantum software system require adequate knowledge and expertise, which play major roles to realise the quantum software design and development activities. This theme is developed to organize the core challenges related to quantum software knowledge and expertise. Following is the detail discussion of the identified challenges (sub-themes).

\textit{Lack of experienced workforce:} Building a workforce for designing a software system is substantially a major challenge in quantum computing domain. The skills needed to develop a classical computing system are different to quantum [S30]. There is a need for specific professional expertise (i.e., human roles in architecture-centric development process) such as quantum software architects, quantum code developers, and quantum domain engineers \cite{khan2022agile}. The technical team should understand physics to characterize the quantum properties of software systems. Such expertise during the quantum software design and architecting phase can enrich the architecting activities to better meet quantum-specific requirements of the software. Designing and architecting quantum software is radically a different concept, and it demands skillful quantum technical and managerial  workforce [S30].  

\textit{Lack of architectural knowledge:} The research field to understand quantum mechanics and integrate it in computing domain by designing quantum software architecture is far from being mature. Various architectural solutions are proposed to develop a quantum software system; however, it require deep knowledge of theory, technology, and understanding to select and implement a suitable solution based on the architectural problem [S31]. It is important to educate the quantum software engineering community to reshape the architecture processes, activities and practices [S31]. 

\begin{tcolorbox}[colback=gray!5!white,colframe=gray!75!black,title=Key Findings of RQ2.5]
\textbf{Finding 14}: Following four core themes of the identified challenges are developed: \textit{quantum data transmission and security, process-centric architecting, architectural tools and technologies}, and \textit{architecting knowledge}.

\textbf{Finding 15}: We observed that most of the (n=6, 40\%) challenges are related to the \textit{architectural tools and technologies} theme. The existing tools and technologies are not at advance level to tackle the architectural problems and it cause various challenges. This is inline with the finding to develop a software engineering community that focuses on devising advance level tools and technologies for managing quantum software architecture challenges \cite{zhao2020quantum}. 
\end{tcolorbox} 

\section{Key Findings and Implications of the SLR} \label{keyfindings}
We now summarise the core findings of the SLR - discussing key results as answers to all RQs - that highlight the state-of-research on architecting quantum software in Section \ref{KeyFindings}. We also discuss the implications of the SLR on future academic research in Section \ref{ResearchImplications} and its significance along with the potential relevance of this SLR to industrial solutions that address challenges of quantum software architecting in Section \ref{IndustrialImplications}.

\subsection{Summary of Key Findings} \label{KeyFindings} 

A conclusive summary of each RQ is presented in Table \ref{tab:KeyFindings} that structures the general demographic details of published research, answering RQ-1.1 to RQ-1.3 and architectural solutions for quantum software answering RQ-2.1 to RQ-2.5. Table \ref{tab:KeyFindings} can be looked up to identify the core finding corresponding to a specific RQ quickly. For example, a summary of the answer to RQ-1.3 highlights that architectural solutions for quantum software can be applied to several domains such as systems and hardware engineering, software engineering, source code compilation, network security, and smart systems.
Since the year 2018, a comparative growth in research on QSE and more specifically quantum software design, architecture, and implementation can be attributed to a number of factors. Our study identifies three such factors including (i) a number of pioneering surveys on quantum software engineering and development \cite{zhao2020quantum}\cite{piattini2021toward}\cite{gill2020quantum}\cite{SE-QC} (ii) community-wide initiatives with dedicated workshops and conferences for quantum software \cite{moguel2020roadmap}\cite{QSEWorkshop}\cite{QSAWorkshop}\cite{barzen2022}, along with the emergence of quantum programming models and languages \cite{zhao2021bugs4q}\cite{garhwal2021quantum}\cite{QuantProg}\cite{OpenQL}. Moreover, beyond academic research, the recently growing interest to exploit quantum computing and technologies in IT industry is based on rapid advances in quantum hardware and quantum programming languages that support QSE initiatives in terms of developing quantum software systems and applications \cite{WinNTMS}\cite{behera2019designing}\cite{courtland2017google}. It is vital to mention that the launch of the Quantum Flagship project in 2018 (funded by the European Commission) reflects regional and global ambitions to foster research and development on quantum computing technologies \cite{QuantumFlagship}.
Similarly, the studies [S31, S34] present solutions that enable software designers and architects to design and implement quantum software using architectural components and connectors that can be mapped to source code modules and interaction between the models. Similarly, Table \ref{tab:KeyFindings} highlights the key findings for RQ-2.3 that to model and represent quantum software architectures the most prominent architectural notations are graph-based modules, box and arrow structures, and Unified Modeling Language. For example, the study [S32] presents a quantum-specific UML named Q-UML that exploits class and sequence diagrams to represent the behavior and structure of quantum software systems. The details in Table \ref{tab:KeyFindings} are self-explanatory and focus on summarising the core findings that have already been discussed in Section \ref{demographyresults} – Section \ref{RQ2}.

\begin{table}[t]
\centering
\scriptsize
\caption{A Summary of the Key Findings of SLR}
\label{tab:KeyFindings}
\begin{tabular}{|ll|}
\hline

\multicolumn{2}{|c|}{\cellcolor[HTML]{EFEFEF}{\color[HTML]{333333} \textbf{Demography of Published Research (RQ-1.1- RQ1.3)}}} \\ \hline

\multicolumn{2}{|l|}{\begin{tabular}[c]{@{}l@{}}\textbf{RQ-1.1} – \textit{Frequency and types of publications}\\ \textbf{Frequency}: Years of publications = 2004 to 2021 with the most number of publications \\from 2018 – 2021 (21, 62\%)\\ 

\textbf{Types}: Journal articles (15, 44\%), Conference proceedings (13, 38\%), Workshop paper (4,  12\%), \\Symposium paper (2,  6\%)\end{tabular}} \\ \hline

\multicolumn{2}{|l|}{\begin{tabular}[c]{@{}l@{}}\textbf{RQ-1.2} – \textit{Types of published research}\\ Personal experience papers, Philosophical papers, Opinion papers, \\Validation research, Proposal of solution, Proposal of solution and validation research\end{tabular}} \\ \hline

\multicolumn{2}{|l|}{\begin{tabular}[c]{@{}l@{}}\textbf{RQ-1.3} – \textit{Application domains of research}\\ Systems and hardware engineering (20, 59\%), Software engineering (5, 15\%),  \\Source code compilation (4, 12\%), \\ Network security (4, 12\%), Smart systems (1, 3\%)\end{tabular}} \\ \hline

\multicolumn{2}{|c|}{\cellcolor[HTML]{EFEFEF}{\color[HTML]{333333} \textbf{Architectural solutions for quantum software (RQ-2.1- RQ-2.5)}}} \\ \hline

\multicolumn{2}{|l|}{\begin{tabular}[c]{@{}l@{}}\textbf{RQ-2.1} – \textit{Architectural process for quantum software}\\ -Architectural requirements\\ -Architectural modeling\\ -Architectural implementation\\ -Architectural validation\\ -Architectural deployment\end{tabular}} \\ \hline

\multicolumn{2}{|l|}{\begin{tabular}[c]{@{}l@{}}\textbf{RQ-2.2} – \textit{Architectural modeling notations}\\ -Graph-based models\\ -Box and arrow\\ -UML\end{tabular}} \\-UML (Q-UML)\\
 \hline

\multicolumn{2}{|l|}{\begin{tabular}[c]{@{}l@{}} \textbf{Q-2.3} – \textit{Architectural patterns}\\ Layered pattern, Pipe and filter pattern, Composite design pattern, Prototype design pattern, \\Recursive containment, Two-qubit gate patterns\end{tabular}} \\ \hline

\multicolumn{2}{|l|}{\begin{tabular}[c]{@{}l@{}}\textbf{Q-2.4} – \textit{Architectural tools and frameworks}\\ XACC, Link layer,  Auto E/E framework, Strawberry fields, qCOR, eQASM, JKQ, Kwant,\\JKQ DDSIM,QuNetSim, GH-QPL\end{tabular}} \\ \hline

\multicolumn{2}{|l|}{\begin{tabular}[c]{@{}l@{}}\textbf{RQ-2.5}  – \textit{Emerging challenges for quantum software architectures}\\ 
-Process-centered architecting\\ -Architecting knowledge and expertise\\ -Quantum data transmission and security\\ -Architectural tools and technologies\end{tabular}} \\ \hline
\end{tabular}
\end{table}

\subsection{Research Implications} \label{ResearchImplications}

\begin{enumerate}[label=(\roman*)]

\item Research types based analysis is performed to understand the types of research conducted by the selected primary studies (see Section \ref{RQ1.2}). However, we found that none of the studies conducted evaluation research to assess a particular problem or solution. Quantum software architecture is an emerging research area and no evaluation research studies conducted to assess the contributions promised by the available architectural solutions. It is a significant research gap, and we encourage the researchers to focus on evaluation research to appraise the real-world significance of quantum software systems as well as the existing relevant architectural problems.
    \item Most of the research studies were conducted across five application areas (see Figure \ref{Fig:Researchfacetsandfocus}(b)); however, we were not able to find enough evidence related to other important areas, like \textit{model-driven quantum software architecture (MDQSA), quantum AI software architecture,} and quantum software architecture applications for the industrial problems \cite{SE-QC}\cite{PISQ}. The possible reason for lack of research in the mentioned areas might be that quantum software architecture is a novel research area and most of the studies focused on proposing architectural solutions for quantum hardware systems (see Figure \ref{Fig:Researchfacetsandfocus} (b)). Therefore, we encourage the research community to put more focus on the following areas: (1) Model driven quantum software architecture (MDQSA) to manage complexity, achieve high level reuse and reduce the development efforts \cite{QSEReport}. (2) Quantum AI software architecture to improve state-of-the-art and propose solutions to operate beyond the classical competencies \cite{graef2021software}. (3) Boost industrial awareness related to quantum software architecture and develop architectural solutions to deal with complex industrial problems \cite{QuantumFlagship}.
    \item Concerning the domain problem, we thoroughly investigated the challenging factors of quantum software architecture (see Section \ref{RQ2.5}) and mapped these factors across different major themes. Thematic mapping provides a conceptual framework to understand the broad picture of the identified challenges and barriers of quantum software architecture \cite{UML}.
\end{enumerate}

In conclusion, this study provides quick access to the body of knowledge based on quantum software architecture literature.

\subsection{Industrial Implications} \label{IndustrialImplications}
\begin{enumerate}[label=(\roman*)]

     \item We systematically investigated, analysed, and mapped the existing tools and frameworks across the architecting activities (see Section \ref{RQ2.4}). A mapping between architecting activities and corresponding tool support can guide practitioners in exploiting the available tool support (enabling automation) to perform a specific architecting activity. For example, as shown in Table \ref{tab:ToolsFramework} if a practitioner wants to conduct architectural validation, he/she can utilise the QuNetSim on implemented architecture to simulate quantum information processing on a quantum network [S25]. In general, the results of this SLR can facilitate the practitioners to get an overview and analyse the extent to which architecting activities, patterns and existing tool support that enable semi-automation can be leveraged to develop industrial-scale solutions for quantum software.
     
     \item We proposed an architecting process, which consists of a sequential list of activities, actions, and events to develop a scalable quantum software architecture (see Section \ref{RQ2.1}). The proposed process acts as a blueprint for practitioners to understand the inputs, workflow, and outputs of the quantum software architecting process (see Section \ref{RQ2.1}). 
     \item Thematic classification of identified challenges (see Figure \ref{Fig:Challengesthemes}) provides an overview of potential barriers that need to consider by practitioners before initiating the architecting activities \cite{QuantProg}. 
    \item Several studies \textit{(n=11, 32\%)} discussed architectural tools and frameworks (see Section \ref{RQ2.4}). We developed a toolchain of the identified tools and frameworks based on their contribution across the architecting activities (see Figure \ref{Fig:Toolsupport}). It will assist the practitioners to select a suitable tool or framework with respect to a specific architecting activity. However, there is still a need for industrial efforts to develop more advanced tools to manage the unexplored architecting activities \cite{OpenQL}\cite{QuantumProspects}.
   
\end{enumerate}
The quantum software architecture is a new and unexplored research area. Academic researchers and industrial practitioners working in quantum software architecture domain are invited to contribute by sharing their experiences. It will alleviate the gap between academic research and industrial practices.

\section{Threats to Validity} \label{threats to validity}
Various threats could impact the validity of this study. However, we adopted the SLR guidelines proposed by Kitchenham and Charters to alleviate these threats \cite{keele2007guidelines}. The potential threats are analyzed based on the core four types of validity threats: internal validity, external validity, construct validity, and conclusion validity \cite{wohlin2012experimentation}\cite{zhou2016map}.

\subsection{Internal validity}
The extent to which certain factors affect the results and analysis of the extracted data is called internal validity. Threats to the internal validity of this study could happen in the following SLR phases:
\paragraph{Search strategy:} It might be possible that relevant primary studies are missed during the search process because of the search strings and the overlap across the selected studies due to the snowballing approach, as highlighted by Jalali and Wohlin \cite{jalali2012systematic}. However, we explicitly defined the search strategy in Section \ref{Search strategy}. The first three authors extracted the search terms based on their understanding of RQs, which were further refined by all the authors in consent meetings. Moreover, the search terms were used to develop the search string, which was iteratively developed by all the authors. It should be noted that the authors have extensive research experience in conducting SLR based studies in the software engineering domain. 
\paragraph{Studies selection and quality assessment:} The inclusion and exclusion criteria are defined in Section \ref{inclusion & exclusions} and used to filter the search results and select the most relevant studies. The first three authors jointly participated in the studies selection process. Furthermore, the first author evaluated the quality of each selected study against the assessment criteria defined in Section \ref{quality assessment}. The second and third authors independently verified the assessment results to avoid personal bias.
\paragraph{Data extraction:} Personal bias is a fundamental data extraction threat in SLR studies.  We mitigate this threat by defining the data extraction form (see Table \ref{Tab:Dataitems}) to consistently extract the relevant data. The first three authors initially extracted the data; however, the other co-authors participated in the discussion meetings to remove any doubt and verify the data as suggested by Wohlin et al. \cite{wohlin2012experimentation}.
\paragraph{Data synthesis:} 
Inaccurate data classification and mapping might cause subjective interpretation bias. However, this threat has been alleviated by following thematic classification guidelines provided by Braun and Clarke \cite{braun2006using}.
Moreover, quantitative and qualitative methods are used to analyze the collected data. The bias in the data synthesis process could impact the data interpretation process. This threat has been lessen by using the well-established descriptive statistical approaches to analyze the quantitative data and thematic mapping for the qualitative data.

\subsection{External validity} External validity refers to the degree to which the study findings could be generalized. We do not claim the generalizability of this study, however; we tried to maximize it by providing an explicit overview of quantum software architecture and logically setting the collected data, results, analysis, and conclusions in the study domain. We followed the rigorous protocol-based SLR approach to attain external validity. Moreover, we followed the guidelines provided by Chen et al. \cite{chen2010towards} to search and select the most appropriate digital repositories and target the relevant peer-reviewed studies. Methodological details (Section \ref{research methodology}, and Figure \ref{Fig:researchmethodologyoverview}), SLR protocol, and data extraction mechanism can support the identification and synthesis of new studies and more RQs to extend this research and minimise the threat to external validity. 

\subsection{Construct validity} A relevant construct validity could be ``data items'' since we as the researches observed, decided, and pick up the text fragments or content from the identified studies. Perhaps, this data extraction might not have been correctly performed due to different reasons. For instance, inappropriate search strategies could cause threats like returning a set of irrelevant studies or missing the relevant articles. We tried to mitigate these threats by following operation measures, e.g., conducting group meetings to finalize the search string, developing studies inclusion and exclusion criteria, performing studies quality assessment, and using data extraction form to remove interpersonal bias. Additionally, the search string is customized according to the peculiarities of the selected databases to identify the most relevant studies.

\subsection{Conclusion validity} Conclusion validity refers to the degree to which the study conclusions are credible  or reasonable. In this SLR, the selection criteria was strict so only quality studies (a clear objective and evaluation) were selected for the analysis in this paper \cite{keele2007guidelines}. Additionally, brainstorming sessions are conducted by the authors to discuss the study findings and draw the correct conclusions. It was acknowledged that beyond the scope of current SLR, future efforts may be needed to evolve the results and conclusions, if newly published research is to be investigated, extending the findings of this SLR.

\section{Related Work} \label{related work}
To the best of our knowledge, this work is the first comprehensive systematic literature review on the research of quantum software architecture, including architecting activities, modelling notations, design patterns, tools and frameworks, and architectural challenges. This section discussed the related work that covers different aspects of quantum software engineering \cite{zhao2020quantum}\cite{gill2020quantum}\cite{garhwal2021quantum}\cite{junior2021systematic}\cite{garcia2022systematic}.

Zhao \cite{zhao2020quantum} conducted a classical survey to cover core quantum software engineering life-cycle activities. Zhao summarised that the quantum software development concept emerges from quantum programming languages, and it is considered synonymous to quantum programming \cite{zhao2020quantum}. However, there is a significant need of complete software engineering discipline for quantum software development. This survey extensively discussed the technological support for quantum software development life-cycle phases, including requirements engineering, design, implementation, testing, and maintenance. The study findings reveal that these areas (phases) are rapidly growing; however, they are still far from being mature.

Gill et al. \cite{gill2020quantum} conducted a comprehensive literature survey to provide in-depth observations of quantum computing concepts and discuss the open challenges experienced by the quantum computing community. A list of taxonomies are proposed to provide conceptual understanding of selecting the available quantum computing techniques and determining the optimal strategies to utilize the classical supercomputing infrastructure. It is because, the existing quantum computers are still not strong enough to replace the supercomputers. Quantum computers are coping with the scaling-up challenge of quantum qubits. It is still not certain when exactly quantum computers will replace the classical; however, it is expected that many exciting improvements will happen in the next decade.

Sunita et al. \cite{garhwal2021quantum} conducted a systematic review that surveys available quantum programming language (QPLs) to overview the state-of-the-art in the context of computer programming for developing quantum-intensive software systems. The study formulates a number of RQs to investigate various aspects of QPL such as types of programming languages, recent trends in the development of QPLs, along with academic and industrial progress on the development and adoption of QPLs. The survey also highlights that well-curated design/architecture for quantum software impacts the selection of QPLs for quantum programming. This SLR is also completed with a recently conducted mixed method research \cite{QuantProg} (mining GitHub repositories and developer survey) to investigate the state-of-practice on QPLs in the context of QSE.

Paulo et al. \cite{junior2021systematic} recently performed a systematic mapping study, reviewing 24 studies, to analyse the existing research on quantum software development in regard to QSE. The focus of the systematic mapping is to understand the prominent programming infrastructures, differences between the development of classical of quantum-intensive software, and the application domain for quantum software systems. The authors highlight that in the last decade the availability of tools, technologies, and programming infrastructures have given impetus to academic on quantum software engineering. 

García et al. \cite{garcia2022systematic} conducted an SLR to explore different types of algorithms developed for quantum machine learning and its applications. The study findings reveal that the various conventional/classical algorithms are used for machine learning solutions in the quantum domain e.g., support vector machine and supervised machines learning k-nearest neighbors (KNN) model. The classical algorithms are mainly used for image classification problems. In broad, the implications of quantum machine learning are promising, however, achieving the full-scale benefits of quantum machine learning is still far from being mature.  The large-scale implications of quantum machine learning algorithms are still exceedingly challenging because of quality, speed and scalability issues. It requires massive improvements in the existing QC infrastructure to tackle complex industrial problems.

\subsection{Comparative analysis}
The comparative analysis of our work with the existing related studies is shown in Table \ref{tab:comparison}. The results reveal that our findings are significantly distinct to the existing related work studies. For instance, sufficient number of primary studies are published, however; only one secondary study \cite{garcia2022systematic} partially followed the formal protocol-based SLR approach to conduct the review study \cite{keele2007guidelines}. The SLR guidelines developed by Kitchenham and Charters are widely adopted to conduct systematic literature reviews in software engineering \cite{keele2007guidelines}. Similarly, we reported the demographic details of each selected primary study, including publication type, frequency, research types, contribution, and application domains (RQ1), which are not considered in the related work secondary studies.

Moreover, we provided a comprehensive overview of quantum software architecture, however Zhao \cite{zhao2020quantum}, and Sunita et al. \cite{garhwal2021quantum} provided introductory level details of quantum software architecture. The subsequent  comparison is made based on architecting activities and modeling notations, which are ignored in the related studies (see Table \ref{tab:comparison}). We developed RQ2.1 and RQ2.2 to respectively define and discuss the key activities of quantum software architecture and modeling notations. Similarly, Zhao \cite{zhao2020quantum} and Paulo et al. \cite{junior2021systematic} provided a simple overview of quantum software design. However, we explicitly cover and discuss the existing design patterns (RQ2.3) used to tackle the commonly occurred quantum software architectural problems.

{\renewcommand{\arraystretch}{1}
\begin{table}[]
\centering
\scriptsize
\caption{A Comparison of Results between this Systematic Review and the Existing Secondary Studies. Note: (\checkmark: included, X: not included,  *: extensive discussion, +: simple overview)}
\label{tab:comparison}
\begin{tabular}{|l|l|l|l|l|l|}
\hline
\multirow{2}{*}{This Review Results} & \multicolumn{5}{c|}{Existing Secondary Studies} \\ \cline{2-6} 
 & \cite{zhao2020quantum} & \cite{gill2020quantum} & \cite{garhwal2021quantum} & \cite{junior2021systematic} & \cite{garcia2022systematic} \\ \hline
Protocol based SLR review & X & X & \checkmark (+) & \checkmark (+) & \checkmark (+) \\ \hline
Demographic detail &X & X & \checkmark (+) & X & X  \\ \hline
Quantum computing basics & \checkmark (*) &\checkmark (*) & \checkmark (+) & \checkmark (+) & \checkmark (+) \\ \hline
Quantum software engineering &\checkmark (*) &\checkmark (+) & X & \checkmark (+) & \checkmark (+) \\\hline
Quantum software architecture & \checkmark (+) & X (+) & \checkmark (+) & X &X \\\hline
Architecture modelling notations & X & X & X & \checkmark (+) & X \\ \hline
Quantum software design patterns & \checkmark (+) & X & X & X & X \\ \hline
Architecture tools and frameworks & X & X & \checkmark (+) & X & X \\ \hline
Challenges & X & X & X & X & X \\ \hline
\end{tabular}
\end{table}}

Additionally, no discussion of quantum software tools and frameworks is provided in the related secondary studies. We explicitly explored the selected primary studies to identify the tools and frameworks that support various architecting activities (RQ2.4). Finally, we reported quantum software architecture challenges and provided their thematic classification map (RQ2.5). However, the existing related studies do not provide any details or abstract level discussion of quantum software architecture challenging factors (see Table \ref{tab:comparison}).

\section{Conclusions} \label{conclusions}
Quantum software architecture -\textit{design and implementation blueprint for quantum software }- represents a new genre of software architectures to address computation-specific challenges rooted in quantum computing. With a growing momentum for the adoption of quantum age systems, industrial initiatives of technology giants (e.g., Google, Microsoft, IBM) and academic research have focused on exploiting architectural solutions to develop quantum software that manages and manipulates quantum hardware. This SLR focused on investigating peer-reviewed published research that streamlines the role of software architectures in designing, implementing, validating, and deploying quantum software. We reviewed a total of 34 qualitatively selected studies to conduct this SLR by answering a total of 08 RQs to a fine-grained presentation of the results. 

Results presents that most of our reviewed studies (n = 21, i.e., 62\% approx.) have been published in the last four years \textit{(2018-2021)}. Majority of the published research types (i.e., proposal of solution and validation research \textit{(n = 20, 59\%)} indicate that quantum software architecture is in its infancy, rapidly evolving by borrowing concepts from classical software architectures to address quantum specific challenges. Quantum-specific challenges include but are not limited to quantum systems co-design and mapping Qubits/Qugates to architectural components and connectors that can be effectively addressed by deriving a process for architecting quantum software. To support the architectural process, modeling notations need to build on established foundations of UML profiles and architectural description languages for (semi-) formal specification of quantum software architectures. The SLR identified a total of five architecting activities, six architectural patterns that promote reuse, 11 tools and frameworks that can automate and customise the process of quantum software architecting. While investigating the architectural challenges, we identified a total of 15 emerging challenging factors, classified across 04 different categories, to resolve emerging issues pertaining to architectural solutions for quantum software. The implications of this SLR are for:
\begin{enumerate}[label=(\roman*)]
\item The researchers interested in focusing on quantum software architecture and willing to fill the open research gaps discussed in the study findings. 
\item Facilitating the knowledge transfer to practitioners regarding quantum software architecture application domains, architecting activities, modeling notations, design patterns, tools and frameworks, and challenges. 
\end{enumerate}

We invite practitioners to step forward to focus more on missing application domains, design architecture description languages, develop tools and frameworks to automate the less focused architecting activities, and propose solutions to tackle the challenging factors. We plan to conduct an empirical study to mine the code hosting and questions and answer public platforms to know practitioners’ perceptions regarding the quantum software architecture. We finally plan to compare the results of the empirical study and this SLR to identify the gap between the research and practice regarding quantum software architecture.
\section*{Acknowledgements}
This work has been supported by the Academy of Finland (project DEQSE 21000055631) and Business Finland (project TORQS 21000057661).

\begin{landscape}
\begin{center}
\textbf{Appendix A}
\end{center}
\footnotesize
\begin{longtable}[c]{|l|p{0.7\linewidth}|c|c|c|}
\caption{Selected studies for this SMS}
\label{tab:selectedstudies}
\\\hline
\textbf{ID} & \textbf{Authors, Publication Title, and   Venue} & \makecell{\textbf{Publication} \\\textbf{Year}} & \makecell{\textbf{Publication} \\\textbf{Type}} & \makecell{\textbf{Quality} \\\textbf{Score}} \\ \hline
\endfirsthead
\endhead
S1 & Vicente Martin, Diego R. López, Alejandro Aguado, Juan Pedro Brito, Julio Setién Villarán, Pedro Jesús Salas Peralta, Carmen Escribano, Víctor Lopez, Antonio Pastor Perales, and Momtchil Peev. \textsf{A Components Based Framework for Quantum Key Distribution Networks.} \textit{In 22nd IEEE International Conference on Transparent Optical Networks (ICTON)}, Bari, Italy, pp.1-4. I, 2020. & 2020 & Conference & 4\\ \hline

S2 & Qiong Li, Dan Le, and Ming Rao. \textsf{A design and implementation of multi-thread quantum key distribution post-processing software}. \textit{In Second IEEE International Conference on Instrumentation, Measurement, Computer, Communication and Control (IMCCC)}, Harbin, China, pp.272-275, 2012. & 2012 & Conference & 3.5 \\ \hline

S3 &Xiang Fu, Leon Riesebos, Lingling Lao, Carmen Garcia Almudever, Fabio Sebastiano, Richard Versluis, Edoardo Charbon, and Koen Bertels. A heterogeneous quantum computer architecture.  In Proceedings of the 13th ACM International Conference on Computing Frontiers, Como, Italy, pp.323-330. 2016.& 2016 & Conference & 5 \\ \hline

S4 & Alexander J.McCaskey, Eugene F.Dumitrescu, Dmitry Liakh, Mengsu Chen, Wu-chun Feng, and Travis S. Humble. A language and hardware independent approach to quantum–classical computing. SoftwareX, 7: pp.245-254, 2018. & 2018 & Journal & 5 \\ \hline

S5 & Krysta M. Svore, Alfred V. Aho, Andrew W. Cross, Isaac Chuang, and Igor L. Markov. A layered software architecture for quantum computing design tools. Computer, 39(1): pp.74-83, 2006.& 2006 & Journal & 4 \\ \hline

S6 & Axel Dahlberg, Matthew Skrzypczyk, Tim Coopmans, Leon Wubben, Filip Rozpędek, Matteo Pompili, Arian Stolk, Przemysław Pawełczak, Robert Knegjens, Julio A De Oliveira Filho, Ronald Hanson,
Stephanie Wehner. A link layer protocol for quantum networks.  In Proceedings of the 33rd ACM Special Interest Group on Data Communication (SIGCOMM), Beijing, China, pp.159-173, 2019.
 & 2019 & Conference & 5\\ \hline

S7 &Thomas Häner, Damian S. Steiger, Krysta Svore, and Matthias Troyer. A software methodology for compiling quantum programs. Quantum Science and Technology , 3(2): pp.1-19, 2018.
& 2018 & Journal & 3.5 \\ \hline

S8 & Michael Booth, Edward Dahl, Mark Furtney, and Steven P. Reinhardt. Abstractions considered helpful: a tools architecture for quantum annealers. In 5th IEEE High Performance Extreme Computing Conference (HPEC), Waltham, MA USA, pp.1-2, 2016. & 2016 & Conference & 2.5 \\ \hline

S9 &Hongbon Lan, Chengrui Zhang, and Hongbin Li. An open design methodology for automotive electrical/electronic system based on quantum platform. Advances in Engineering Software, 39 (6): pp. 526-534, 2008. & 2008 & Journal & 3 \\ \hline

S10 & Victor Potapov, Sergei Gushansky, Vyacheslav Guzik, and Maxim Polenov. Architecture and software implementation of a quantum computer model.  In 2nd Computer Science On-line Conference (CSOC), pp. 59-68. Springer, Cham, Zlin, Czech Republic, pp.59-68, 2016. & 2016 & Conference & 4 \\ \hline

S11 & Loyd R. Hook, and Samuel C. Lee. Design and simulation of 2-D 2-dot quantum-dot cellular automata logic. IEEE Transactions on Nanotechnology, 10(5), pp.996-1003, 2010. & 2010 & Journal & 5\\ \hline

S12 & Ilia Polian, and Austin Fowler. Design automation challenges for scalable quantum architectures. In 52nd ACM/EDAC/IEEE Design Automation Conference (DAC), Austin, TX, USA, pp.1-6, 2015. & 2015 & Conference & 3.5\\ \hline

S13 &Heranmoy Maity,  Arijit Kumar Barik, Arindam Biswas, Anup Kumar Bhattacharjee, and Anita Pal. Design of quantum cost, garbage output and delay optimized BCD to excess-3 and 2’s complement code converter. Journal of Circuits, Systems and Computers, 27(12): pp.1-5, 2018. & 2018 & Journal & 4\\ \hline

S14 & Xiang Fu, Leon Riesebos, Adriaan Rol, Jeroen Van Straten, Hans van Someren, Nader Khammassi, Imran Ashraf, Raymond Vermeulen, V. Newsum, Kelvin Kwong Lam Loh, Jacob de Sterke, Wouter Vlothuizen, Raymond Schouten, Carmen G. Almudéver, Leonardo DiCarlo, and Koen Bertels. eQASM: An executable quantum instruction set architecture. In 25th IEEE International Symposium on High Performance Computer Architecture (HPCA), Washington, DC, USA, pp.224-237, 2019. & 2019 & Symposium & 4.5\\ \hline

S15 &Prakash Murali, Norbert Matthias Linke, Margaret Martonosi, Ali Javadi Abhari, Nhung Hong Nguyen, and Cinthia Huerta Alderete. Full-stack, real-system quantum computer studies: Architectural comparisons and design insights. In 46th ACM/IEEE  Annual International Symposium on Computer Architecture (ISCA), Phoenix, AZ, USA, pp.527-540. 2019. & 2019 & Conference & 5\\ \hline

S16 & Robert Wille, Stefan Hillmich, and Lukas Burgholzer. JKQ: JKU tools for quantum computing. In 33rd IEEE/ACM International Conference On Computer Aided Design (ICCAD), San Diego, CA, USA, pp.1-5, 2020.& 2020 & Conference & 4 \\ \hline

S17 & Christoph W Groth, Michael Wimmer, Anton R. Akhmerov, and Xavier Waintal. Kwant: a software package for quantum transport. New Journal of Physics, 16 (6): pp.1-40, 2014. & 2014 & Journal & 4.5\\ \hline

S18 & Nathan Cody Jones, Rodney Van Meter, Austin Fowler, Peter McMahon, Jungsang Kim, Thaddeus Ladd, and Yoshihisa Yamamoto. Layered architecture for quantum computing. Physical Review X, 2(3): pp.1-27, 2012. & 2012 & Journal & 5\\ \hline

S19 & Alwin Zulehner, and Robert Wille. Advanced simulation of quantum computations. IEEE Transactions on Computer-Aided Design of Integrated Circuits and Systems, 38 (5): pp. 848-859, 2018.  & 2018 & Journal & 5\\ \hline

S20 &Gushu Li, Anbang Wu, Yunong Shi, Ali Javadi-Abhari, Yufei Ding, and Yuan Xie. On the Co-Design of Quantum Software and Hardware. In Proceedings of the 8th Annual ACM International Conference on Nanoscale Computing and Communication (NANOCOM), Italy, pp.1-7, 2021. & 2021 & Conference & 3.5 \\ \hline

S21 &Teague Tomesh, and Margaret Martonosi. Quantum Codesign. IEEE Micro, 41(5), pp-33-40, 2021.  & 2021 & Journal & 3.5\\ \hline

S22 & Munish Bhatia, and Avneet Kaur. Quantum computing inspired framework of student performance assessment in smart classroom. Transactions on Emerging Telecommunications Technologies, 32(9): pp.1-22, 2021. & 2021 & Journal & 3.5\\ \hline

S23 & Nan Wu, Haixing Hu, Fangmin Song, Huimin Zheng, and Xiangdong Li. Quantum software framework: a tentative study. Frontiers of Computer Science, 7(3): pp.341-349, 2013. & 2013 & Journal & 5\\ \hline

S24 & Iaakov Exman, and Alon Tsalik Shmilovich. Quantum Software Models: The Density Matrix for Classical and Quantum Software Systems Design. In Proceedings of the IEEE/ACM 43rd International Conference on Software Engineering Workshops (ICSEW), Madrid, Spain, pp.1-6, 2021. & 2021 & Workshop & 5\\ \hline

S25 & Stephen Diadamo, Janis Nötzel, Benjamin Zanger, and Mehmet Mert Beşe. Qunetsim: A software framework for quantum networks. IEEE Transactions on Quantum Engineering, 2: pp.1-12, 2021. & 2021 & Journal & 3.5\\ \hline

S26 & Lalitha Nallamothula. Selection of quantum computing architecture using a decision tree approach. In 3rd International Conference on Intelligent Sustainable Systems (ICISS), Thoothukudi, India, pp.644-649, 2020. & 2020 & Conference & 5\\ \hline

S27 & Killoran, Nathan, Josh Izaac, Nicolás Quesada, Ville Bergholm, Matthew Amy, and Christian Weedbrook. . Strawberry fields: A software platform for photonic quantum computing. Quantum, 3: pp-1-27, 2019. & 2019 & Journal & 3.5 \\ \hline

S28 &Alexander Mccaskey, Thien Nguyen, Anthony Santana, Daniel Claudino, Tyler Kharazi, and Hal Finkel. Extending c++ for heterogeneous quantum-classical computing. ACM Transactions on Quantum Computing, 2( 2):pp. 1-36, 2021. & 2021 & Journal & 5\\ \hline

S29 & Krista  Svore, Andrew Cross, Alfred Aho, Isaac Chuang, and Igor Markov. Toward a software architecture for quantum computing design tools. In Proceedings of the 2nd International Workshop on Quantum Programming Languages (QPL), pp. 145-162. 2004.& 2004 & Workshop & 3.5\\ \hline

S30 & Frank Leymann. Towards a pattern language for quantum algorithms. In International Workshop on Quantum Technology and Optimization Problems (QTOP), Springer, Cham, Munich, Germany, pp. 218-230, 2019. & 2019 & Workshop & 3.5\\ \hline

S31 & Frank Leymann, Johanna Barzen, and Michael Falkenthal. Towards a platform for sharing quantum software. Proceedings of the 13th Advanced Summer School on Service Oriented Computing (SummerSOC), Crete, Greece,  pp.70-74, 2021. & 2021 & Conference & 3\\ \hline

S32 & Carlos A. Pérez-Delgado, and Hector G. Perez-Gonzalez. Towards a quantum software modeling language. In Proceedings of the IEEE/ACM 42nd International Conference on Software Engineering Workshops (ICSEW), Seoul, South Korea, pp.442-444, 2020. & 2020 & Workshop & 2.5\\ \hline

S33 & El-Mahdy M.Ameen, , Hesham A. Ali, Mofreh M. Salem, and Mahmoud Badawy. Towards implementation of a generalized architecture for high-level quantum programming language. International Journal of Theoretical Physics, 56(8): pp.2376-2412, 2017. & 2017 & Journal & 3.5 \\ \hline

S34 & Rob F.M. van den Brink, Frank Phillipson
, and Niels M.P. Neumann. Vision on next level quantum software tooling. In Proceedings of the 10th International Conference on Computational Logics, Algebras, Programming, Tools, and Benchmarking (COMPUTATION TOOLS), Venice, Italy, pp.16-23, 2019.
 & 2019 & Conference & 5 \\ \hline

\end{longtable}

\end{landscape}

\begin{table}[h]
\centering
\caption{Cohen’s Kappa test R-Code}
\label{tab:Cohens Kappa test R-Code}
\begin{tabular}{|l|}
\hline
\rowcolor[HTML]{EFEFEF} 
\multicolumn{1}{|c|}{ 
\textbf{Cohen’s Kappa test between authors for the SLR process}}         \\ \hline
{ \begin{tabular}[c]{@{}l@{}}library(DescTools)\\ \\ QuantumSoftwareArch \textless{}- data.frame(Authors1to3=c(7,6,4,3,10,2,9,5,1,3),\\ \\ Authors4to5=c(7,6,4,3,9,2,8,5,1,3),\\ \\ Authors6to7=c(7,5,3,2,10,2,9,5,1,3))\\ \\ KappaM(QuantumSoftwareArch)\\ \\ KappaM(QuantumSoftwareArch, method="Conger")\\ \\ KappaM(QuantumSoftwareArch, conf.level=0.95)\\ \\ KappaM(QuantumSoftwareArch, method="Light")\end{tabular}} \\ \hline
 
\multicolumn{1}{|c|}{ \textbf{Cohen’s Kappa test between authors for the snowballing process}} \\ \hline
{ \begin{tabular}[c]{@{}l@{}}library(DescTools)\\ \\ Snowballing \textless{}- data.frame(Authors1to4=c(1,4,2,3,5),\\ \\ Authors5to6=c(1,5,2,3,4))\\ \\ KappaM(Snowballing)\\ \\ KappaM(Snowballing, method="Conger")\\ \\ KappaM(Snowballing, conf.level=0.95)\\ \\ KappaM(Snowballing, method="Light")\end{tabular}} \\ \hline
\end{tabular}
\end{table}
\clearpage


\clearpage
\bibliographystyle{unsrt}
\bibliography{References}

\end{document}